\documentclass[aps,prc,twocolumn,superscriptaddress,showpacs,preprintnumbers,amsmath,amssymb]{revtex4}
\usepackage{graphicx}% Include figure files
\usepackage{xcolor}%

\begin{document}

%\vspace*{1.5cm}
\def	\evistot	{E_{\rm vis}^{\rm tot}}
\def	\evis	{E_{\rm vis}}
\def	\nhit	{N_{\rm hit}}
\def	\snrad	{\rm^{107}Sn}
\def	\sn	{\rm^{124}Sn}
\def	\la	{\rm^{124}La}
\def	\target	{\rm^{nat}Sn}
\def	\zbound	{Z_{\rm bound}}
\def	\pxmin	{p_{x1}}
\def	\pxmax	{p_{x2}}
\def	\pymin	{p_{y1}}
\def	\pymax	{p_{y2}}
\def	\ave	{``AVE''}
\def	\min	{``MIN''}
\def	\max	{``MAX''}
\def	\Nwin	{N^{\rm win}}
\def	\Ns	{N_{\rm S}}
\def	\Nb	{N_{\rm B}}
\def	\Nswin	{N_{\rm S}^{\rm win}}
\def	\Nbwin	{N_{\rm B}^{\rm win}}
\def	\Nobs	{N_{\rm obs}^{\rm win}}
\def	\Gt	{G_{T}}	
\def	\ktot	{k_{\rm tot}}
\def	\Yexp	{Y_{\rm exp}}
\def	\Ypois	{Y_{\rm Pois}}
\def	\Yf	{Y_{\rm f}}
\def	\Cf	{C_{\rm f}}

%% Definitions of units %%
\def	\ns	{{\rm ns}}
\def	\cm	{ {\rm cm}}
\def	\MeV	{ {\rm MeV}}
\def	\AMeV	{ {\rm AMeV}}
\def	\MeVc	{ {\rm MeV/c}}
\def	\AMeVc	{ {\rm AMeV/c}}

%% definitions of numeric constants %%
\def	\elab	{E_{\rm LAB}=600\ \AMeV}	%LAB energy
\def	\eps	{35_{-5}^{+10}\ \MeV}	%Total Evis/N
\def	\projmom	{1215\ \AMeVc}		%Projectile momentum
\def	\eff	{0.73}			%LAND efficiency
\def	\valpxmin	{-150\ \MeVc}		%px_min
\def	\valpxmax{0}			%px_max
\def	\valpymin	{-50\ \MeVc}		%py_min
\def	\valpymax{50\ \MeVc}		%py_max

\def\GSI{GSI Helmholtzzentrum f\"{u}r Schwerionenforschung GmbH, D-64291 Darmstadt,
Germany}
\def\GOETHE{Institute for Theoretical Physics, J. W. Goethe University, D-60438 Frankfurt am Main, Germany}
\def\HFHF{Helmholtz Research Academy Hesse for FAIR (HFHF), Max-von-Laue-Str. 12, D-60438 Frankfurt am Main, Germany}
\def\KONYA{Department of Physics, University of Sel\c{c}uk, TR-42079 Konya, Turkey}
\def\CHALMERS{Chalmers University of Technology, Kemiv\"{a}gen 9, S-412 96 G\"{o}teborg, Sweden}
\def\FIAS{Frankfurt Institute for Advanced Studies, J.W. Goethe University, D-60438 Frankfurt am Main, Germany}
\def\JAGU{M. Smoluchowski Institute of Physics, Jagiellonian University, PL-30059 Krak\'{o}w, Poland}
\def\LPC{LPC, IN2P3-CNRS, ISMRA et Universit{\'e}, F-14050 Caen, France}
\def\GANIL{GANIL, CEA et IN2P3-CNRS, F-14076 Caen, France}
\def\IFJ{H. Niewodnicza{\'n}ski Institute of Nuclear Physics,
PL-31342 Krak{\'o}w,
Poland}
\def\CATANIA{Dipartimento di Fisica e Astronomia-Universit\`a and INFN-Sezione CT and LNS, 
I-95123 Catania, Italy}
\def\WARSAW{National Centre for Nuclear Research, PL-02093 Warsaw, Poland}
\def\MAINZ{Institute of Nuclear Physics, Johannes Gutenberg University, D-55099 Mainz, Germany}

\newcommand{\goo}{\,\raisebox{-.5ex}{$\stackrel{>}{\scriptstyle\sim}$}\,}
\newcommand{\loo}{\,\raisebox{-.5ex}{$\stackrel{<}{\scriptstyle\sim}$}\,}

\title{Neutrons from projectile fragmentation at 600 MeV/nucleon}

\affiliation{\IFJ}
\affiliation{\JAGU}
\affiliation{\KONYA}
\affiliation{\CHALMERS}
\affiliation{\GSI}
\affiliation{\WARSAW}
\affiliation{\GOETHE}
\affiliation{\HFHF}
\affiliation{\GANIL}
\affiliation{\FIAS}
\affiliation{\MAINZ}

\author{P.~Paw{\l}owski}        \affiliation{\IFJ}
\author{J.~Brzychczyk}          \affiliation{\JAGU}   
\author{N.~Buyukcizmeci}    	\affiliation{\KONYA}
\author{H.~T.~Johansson}        \affiliation{\CHALMERS} 
\author{W.~Trautmann}           \affiliation{\GSI}
\author{A.~Wieloch}             \affiliation{\JAGU} 
\author{P.~Adrich}              \affiliation{\WARSAW}  
\author{T.~Aumann}              \affiliation{\GSI} 
\author{T.~Barczyk}\thanks{deceased}             \affiliation{\JAGU} 
\author{S.~Bianchin}            \affiliation{\GSI} 
\author{K.~Boretzky}            \affiliation{\GSI}  
\author{A.~S.~Botvina}       	\affiliation{\GOETHE}\affiliation{\HFHF}%\affiliation{\MOSCOW}
\author{A.~Chbihi}              \affiliation{\GANIL}
\author{J.~Cibor}               \affiliation{\IFJ}
\author{B.~Czech}               \affiliation{\IFJ} 
\author{H.~Emling}              \affiliation{\GSI}
\author{J.~D.~Frankland}           \affiliation{\GANIL}
\author{M.~Heil}                \affiliation{\GSI}
\author{A.~Le~F\`evre}          \affiliation{\GSI}
\author{Y.~Leifels}             \affiliation{\GSI}
\author{J.~L\"{u}hning}         \affiliation{\GSI}
\author{J.~{\L}ukasik}          \affiliation{\IFJ}\affiliation{\GSI} 
\author{U.~Lynen}               \affiliation{\GSI} 
\author{Z.~Majka}               \affiliation{\JAGU}  
\author{I.~N.~Mishustin}     \affiliation{\FIAS}%\affiliation{\KURCH}
\author{W.~F.~J.~M\"{u}ller}      \affiliation{\GSI}
\author{R.~Ogul}        	\affiliation{\KONYA}
\author{H.~Orth}                \affiliation{\GSI}
\author{R.~Palit}               \affiliation{\GSI}
\author{D.~Rossi}               \affiliation{\GSI}
\author{C.~Schwarz}             \affiliation{\GSI}
\author{C.~Sfienti}             \affiliation{\MAINZ}
\author{H.~Simon}               \affiliation{\GSI}
\author{K.~S\"{u}mmerer}        \affiliation{\GSI}
\author{H.~Weick}               \affiliation{\GSI}
\author{B.~Zwiegli\'{n}ski}     \affiliation{\WARSAW}

\date{\today}% It is always \today, 

\begin{abstract}
The neutron emission in projectile fragmentation at relativistic energies was
studied with the Large-Area-Neutron-Detector LAND coupled to the ALADIN forward spectrometer at 
the GSI Schwerionen-Synchrotron (SIS). 
Stable $^{124}$Sn and radioactive $^{107}$Sn and $^{124}$La beams with an incident energy of 
600 MeV/nucleon were used to explore the $N/Z$ dependence of the identified neutron
source. 
A cluster-recognition algorithm is applied for identifying individual particles within the 
hit distributions registered with LAND.
The obtained momentum distributions are extrapolated over the full phase space occupied
by the neutrons from the projectile-spectator source. 
The mean multiplicities of spectator neutrons reach values of up to about 11 and depend 
strongly on the isotopic composition of the projectile. An effective source temperature of
$T \approx 2 - 5$~MeV, monotonically increasing with decreasing impact parameter,
is deduced from the transverse momentum distributions. 
For the interpretation of the data, calculations with the statistical multifragmentation 
model were performed. The variety of excited projectile spectators assumed to decay statistically
is represented by an ensemble of excited sources with parameters determined previously from
the fragment production observed in the same experiments.
The obtained agreement is very satisfactory for more peripheral collisions where, according to the model, 
neutrons are mainly emitted during the secondary decays of excited fragments. 
The neutron multiplicity in more central collisions is underestimated, indicating that other sources besides
the modeled statistical breakup contribute to the observed neutron yield. 
The choice made for the symmetry-term coefficient of the liquid-drop description of produced fragments has a weak effect on the
predicted neutron multiplicities. 
\end{abstract}


\maketitle

\section{INTRODUCTION}
\label{sec:int}

A comprehensive study of the isospin 
dependence of projectile fragmentation at relativistic energies has been performed at the GSI Schwerionen-Synchrotron (SIS) 
with stable and radioactive Sn and La beams of 600 MeV/nucleon~\cite{sfienti09,ogul11}. 
As a primary result, it was found that the fragmentation process as manifested by the recorded fragment distributions
and correlations responds only weakly to changes in the projectile composition explored within the interval
of neutron-to-proton ratios $N/Z = 1.16 - 1.48$. Among global observables, only the isotopic fragment distributions, 
measured with individual mass resolution up to atomic number $Z \le 10$, varied as a function of the projectile $N/Z$.
Chemical breakup temperatures, as determined from double yield ratios of $Z = 2 - 4$ isotopes, were in the range of
$T = 4 - 8$~MeV, increasing with decreasing impact parameter, and found to not depend on the isotopic composition of the projectile~\cite{sfienti09}. 
Temperatures of $T \approx 6$~MeV characterize the class of reactions associated with a maximum production of intermediate-mass fragments.

One of the main observations concerned the abundance of neutron-rich fragments in the reaction
channels leading to the multifragmentation of the excited projectile-spectator systems.
In particular, in the range of fragments with atomic numbers $6 \le Z \le 10$, the neutron
richness of the measured mass distributions exceeded the predictions of the statistical
multifragmentation model (SMM, Ref.~\cite{smm}) whose input parameters had been adjusted to
reproduce the measured fragment $Z$ distributions and correlations. For also correctly reproducing the 
mass distributions, a significant reduction of the symmetry-term coefficient in the
liquid-drop description of the produced fragments was found to be necessary. This was supported 
by the isoscaling analysis performed with the same data~\cite{ogul11} which confirmed earlier
findings for similar reactions~\cite{LeFevre05,henzlova10}.

\begin{figure*} 
%\centerline{\includegraphics[width=7.5cm,angle=-90]{neutrons_fig1.eps}}
\centerline{\includegraphics[width=17.0cm]{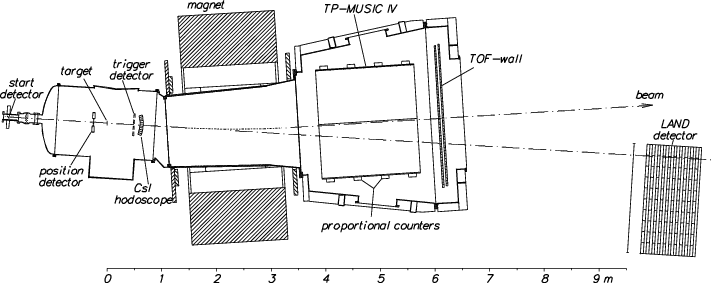}}
%\centerline{\includegraphics[width=16.5cm]{../tt_pict/s254_tt_prag_thicker.eps}}
\caption{\small{Setup of the S254 experiment consisting of the ALADIN forward spectrometer and the neutron detector LAND in a 
cross-sectional view from above onto the plane containing the beam axis.
The beam enters from the left. The white rectangles in the magnet indicate the location of the coils. The instrumentation installed in the target
chamber is described in Ref.~\protect\cite{ogul11}; 
data from the 84-element CsI hodoscope~\protect\cite{kunde95} were not used for the present analysis
(reprinted from Ref.~\protect\cite{sfienti05}, Copyright $\copyright$ 2005, 
with permission from Elsevier).
}}
\label{fig:setup}
\end{figure*}  %fig. 1

Apart from the fast nucleons emitted during the primary cascade-type stage of the reaction,
neutron emission is expected to occur primarily in the form of secondary decays of the excited residue
nuclei and fragments. 
The observed neutron richness of the fragments
implies that fewer neutrons are evaporated than one would expect with standard SMM assumptions. 
The properties of the neutron emission should thus complement the observations made 
in the fragment channels. 

In this paper, we report the results for the projectile-spectator source of neutrons obtained in the same experiments. 
This was made possible by positioning the
Large-Area Neutron Detector LAND~\cite{LAND} downstream of the ALADIN forward spectrometer.
The acceptance of LAND covered approximately half of the phase space occupied by neutrons from the projectile decay
which was sufficient for identifying the strength and main dynamical properties of the projectile neutron source. 
For the theoretical interpretation of the experimental findings, SMM  
calculations~\cite{botvina95} were performed. 
The same ensembles and parameters were used that have previously been determined in the study 
of fragment production in these reactions~\cite{ogul11}.

LAND had already been coupled to the ALADIN spectrometer in earlier experiments. The 
properties of neutron emission obtained in this way have been essential for determining the
average energy transfer to the projectile spectator as a function of the impact 
parameter~\cite{poch95,odeh00}. The separation energies of neutrons and their kinetic 
energies in the source frame amount together to a substantial fraction of the total
excitation energies of the produced spectator sources (see, e.g., Fig.~7 in 
Ref.~\cite{viola_wci06}). 

Multifragmentation experiments including the measurement of neutrons, even though difficult 
because of the different techniques required for simultaneously detecting neutral and 
charged particles, have been performed at several 
laboratories (see, e.g., Refs.~\cite{jahnke83,galin94,kunde96,sobot00,wuenschel09,morfouace19} and references given therein). 
Neutron emission in heavy-ion
reactions at relativistic energies has been studied with plastic scintillator arrays at 
the Bevalac by the group of Madey {\it et al.} in inclusive~\cite{madey81,madey85,baldwin92} and
impact-parameter selected measurements~\cite{madey88,htun99}. Together with the present results, 
they reveal common properties of the neutron emission in spectator fragmentation, as will be shown in the discussion sections.

Results of the ALADIN-LAND experimental study have been used in a variety of different analyses. 
Besides the statistical description obtained with the SMM and reported in Ref.~\cite{ogul11}, 
a dynamical description of the observed fragment production was achieved by Su {\it et al.}~\cite{junsu18}. 
It is based on calculations using the isospin dependent
quantum molecular dynamics (IQMD) model and a minimum spanning tree algorithm for recognizing fragments after their formation. 
The statistical code GEMINI was used to investigate the influence of secondary decays on fragmentation observables. 

Very recently, fluctuations up to fourth order of the $Z$ distributions of the largest fragment in these reactions were 
investigated with the experimental data as well as statistical calculations with the SMM and the canonical thermodynamic 
fragmentation model~\cite{pietrzak20}. 
They were found to exhibit signatures characteristic of a second-order phase transition, 
established with cubic bond percolation and previously observed for the fragmentation of $^{197}$Au projectiles 
at similar energies~\cite{jb18}.
These signatures depend only weakly on the $A/Z$ ratio of the fragmenting spectator source. 
The transition point is characterized by the asymmetry parameter skewness passing through zero and the kurtosis excess simultaneously
reaching a minimum. 
According to the IQMD calculations for these reactions, it is located at fairly 
peripheral impact parameters $b = 8.5 - 9.0$~fm~\cite{junsu22,xiao22}. 

The minor influence of 
the isotopic composition on general properties of the fragmentation process has emerged as a common observation 
in all studies made so far~\cite{sfienti09,ogul11,junsu18,pietrzak20}. The only exception is the mean neutron-over-proton 
ratio of intermediate mass fragments found to depend on the neutron richness of the initial projectiles in the present 
and similar reactions~\cite{ogul11,henzlova08,foehr11,imal15}. Also for that reason, the properties 
of the neutron source are of interest and can be expected to provide complementary information.

%insert role in colliders as discussed in 
%Svetlichnyi, A.; Nepeivoda, R.; Pshenichnov, 
%I. Using Spectator Matter for Centrality Determination in Nucleus-Nucleus Collisions.
%Particles 2021, 4, 227–235. 
%https://doi.org/10.3390/particles4020021
%cite nima_906 2018 114_pshenichnov_arx_1805.01792v2

\section{EXPERIMENTAL DETAILS}
\label{sec:exp}

\subsection{ALADIN experiment S254} 
\label{sec:s254}

%ALADIN + running conditions (from Ogul {\it et al.})
The ALADIN experiment S254, conducted in 2003 at the SIS heavy-ion synchrotron,
has been described in detail in the previous Refs.~\cite{sfienti09,ogul11}.
Stable $^{124}$Sn and neutron-poor secondary $^{107}$Sn and $^{124}$La beams were used 
for the study of isotopic and isobaric effects over a wide range of isotopic 
compositions. The radioactive secondary beams were produced at the fragment 
separator FRS~\cite{frs92} by fragmenting primary $^{142}$Nd 
projectiles in a thick beryllium  target. To reach the necessary intensities,
contributions of neighboring nuclides in addition to the requested isotopes
had to be accepted. The mean compositions of the nominal $^{124}$La ($^{107}$Sn) beams 
were $\langle Z \rangle$ = 56.8 (49.7) and mass-over-charge ratio $\langle A/Z \rangle$ = 2.19 (2.16), 
respectively~\cite{luk08}. 
All beams had a laboratory energy of 600 MeV/nucleon, intensities of about 10$^3$ 
particles/s, and were directed onto
reaction targets consisting of $^{\rm nat}$Sn with areal density 500 or 1000 mg/cm$^2$,
corresponding to estimated interaction probabilities slightly exceeding 1\% and 2\%, respectively.

The ALADIN experimental setup has been described previously~\cite{schuett96}. Its 
configuration in the present experiment is shown in Fig.~\ref{fig:setup}
in the form of a cross-sectional view from above onto the horizontal plane containing the beam axis.
It included three chambers: the target chamber followed by the chamber inserted in between the
poles and coils of the window-frame type magnet and the detector chamber. Vacuum is maintained in the target and 
magnet chambers up to the pressure window separating the volumes of the magnet and detector chambers. 
The gas pressure in the detector chamber was kept at approximately 1 bar.

Fragment detection and identification was achieved by measuring the atomic number $Z$,
the magnetic rigidity, and the velocity of the produced fragments with the 
TP-MUSIC IV detector, upgraded by adding proportional counters to the readout system~\cite{bauer97,sann97,sfienti03},
and the ALADIN time-of-flight wall~\cite{hubele91}.
The threshold for fragment detection and identification was below $Z=2$. The
obtained resolution in atomic number was $\Delta Z \leq 0.6$ (FWHM), independent of $Z$ 
up to the projectile $Z$. Fragment masses were identified with a resolution of about 3\% for 
fragments with $Z \le 3$ (standard deviation), improving to 1.5\% for $Z\geq 6$, and found 
to be the same in the experiments with stable and radioactive beams. Masses are thus 
individually resolved for fragments with $Z \leq 10$.

The acceptance of the ALADIN forward spectrometer, in the geometry of the present experiment,
was $\pm 10.2^{\circ}$ for $N=Z$ fragments with beam velocity in the horizontal direction,
i.e., in the bending plane, and $\pm 4.5^{\circ}$ in the vertical direction. The magnetic fields 
were set to deflect the primary ion beams by an angle of 7.1$^{\circ}$. The neutron detector
LAND was positioned approximately 10 m downstream from the target, in the vertical direction symmetrically with respect 
to the incoming beam direction but displaced horizontally away from the direction of the
deflected beam (Fig.~\ref{fig:setup}). The intersection point of the original direction of the incoming beam with 
the front face of the veto wall of LAND had a distance of 9.65 m from the target and of 22 cm 
from the left edge of the leftmost paddle of the wall. At this distance, the acceptance of LAND 
for neutrons emitted from the target position is unobstructed by the pole gap of the ALADIN magnet, except for 
small regions near the upper, lower, and right edges of the detector. Roughly one half of the 
phase space occupied by the spectator source of neutrons is covered with LAND in this 
configuration.

Trigger signals for reactions in the target were derived from four plastic-scintillator paddles
positioned approximately 50 cm downstream from the target at angles outside the acceptance of 
the spectrometer. The condition that at least one of them fired was met with nearly 100\% 
probability by events with moderate to large charged-particle multiplicities~\cite{ogul11}. 

The quantity $Z_{\rm bound}$ defined as the sum of the atomic numbers $Z_i$ of all detected
fragments with $Z_i \geq$ 2 was chosen as the principal variable for event sorting. 
Because of the selective coverage of the projectile-spectator decay, $Z_{\rm bound}$ 
represents approximately the charge of the primary spectator system, 
apart from emitted hydrogen isotopes, and is monotonically correlated with the impact 
parameter of the reaction \cite{xiao22,schuett96,hubele91}. 
The excitation energy per nucleon of the spectator system is 
inversely correlated with $Z_{\rm bound}$~\cite{schuett96,ogilvie91}.

The measurement of neutrons with LAND coupled to the ALADIN spectrometer is characterized by two specific features. 
One is the use of LAND for reactions producing large neutron multiplicities, which has the effect that the identification of 
individual neutrons from the hit pattern recorded with the detector is not straightforward.
This topic was investigated prior to the final data analysis, and the obtained results are published in Ref.~\cite{piotr_nima}.
The second distinctive item concerns the detection of neutrons originating from the target before the ALADIN spectrometer with considerable 
amounts of material, mostly iron, being placed in the area between the target and the neutron detector (Fig.~\ref{fig:setup}). 
The feasability of a reliable measurement under these conditions was explored with a model study
using the Geant4 framework~\cite{geant4} which is reported in detail in the Appendix of this paper. 
The essential results of both studies and their consequences for the data analysis and achievable accuracy of the measurement are 
summarized in the following subsections.

\subsection{LAND detector} 
\label{sec:land}

\begin{figure} [htb]
\centerline{\includegraphics[width=7.0cm,keepaspectratio]{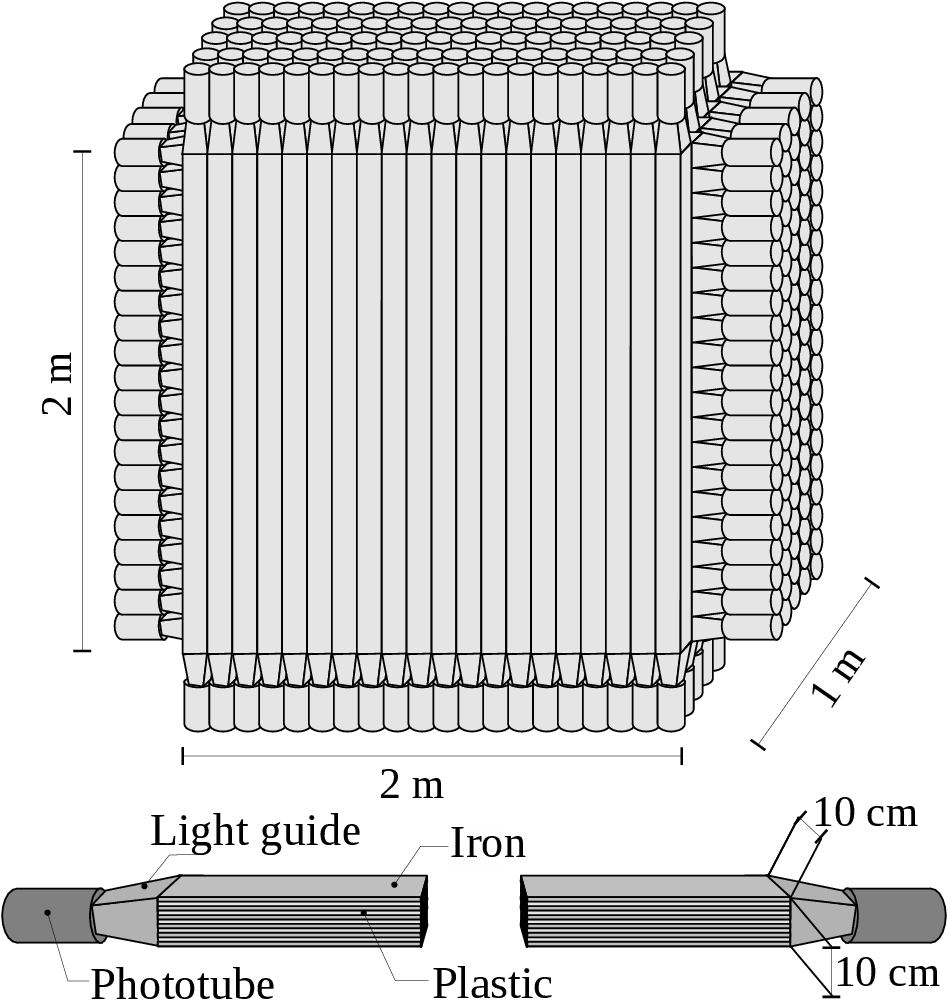}}
%\centerline{\includegraphics[width=8cm,keepaspectratio]{figures_wieloch/iwm2007_proc_fig1_new_v3.eps}}
\caption{\small{Schematic views of the LAND detector without the veto wall and of the internal
structure of a paddle.}}
\label{fig:land}
\end{figure}

The Large-Area-Neutron-Detector LAND is a $2 \times 2 \times 1$ m$^3$ calorimeter consisting of in total 200 slabs
of interleaved iron and plastic strips viewed by photomultiplier tubes at both ends~\cite{LAND}.
A $2 \times 2$~m$^2$ veto wall in front of the detector permitted the identification of light charged particles, 
mainly hydrogen isotopes, hitting the LAND detector. The wall consists 
of a plane of 20 vertical paddles made of 5-mm-thick plastic scintillators, each 2 m long
and 10 cm wide. 
Their area and orientation are the same as those of the first plane of LAND (Fig.~\ref{fig:land}).

   %fig. 2

The main part of LAND consists of ten consecutive planes, each plane containing twenty 
detector slabs, so-called paddles, with a length of 2 m and a cross section of $10 \times 10$ cm$^2$.
The orientation of the paddles is alternating between vertical and horizontal, 
i.e. different for each pair of neighboring planes (Fig.~\ref{fig:land}).
The paddles are made of alternating layers of plastic scintillator and iron converter sheets, all 5 mm
thick, except for the front and rear sheets of iron which are 2.5 mm thick. 
Photo-multiplier tubes mounted at both ends collect the light produced in all plastic layers of a paddle. 
The hit position within a paddle is determined from the ratio of amplitudes or from
the time difference of the two signals recorded at either end. The intrinsic resolutions are 250~ps for the time and 
3~cm for the position (rms values, Ref.~\cite{yordanov05}).

Global observables associated with a reaction event are generated by counting the total
number of hits observed, called hit multiplicity, and by integrating the recorded pulse 
heights to obtain the total visible energy $E_{\rm tot}^{\rm vis}$. By using calibration parameters deduced from
recorded cosmic-ray events, the visible energy can be expressed in MeV~\cite{LAND,piotr_nima}.
Because of its depth of 1 m, the efficiency of LAND for detecting neutrons is high. According to the calibration obtained 
from the S107 experiment performed in 1992 with the fully functioning detector, neutrons of 600 MeV are detected with
efficiency 0.94~\cite{boretzky03,gsi1992}, a value in good agreement with recent Geant4 calculations~\cite{boretzky21}.
%The losses of 6\% combine a small number of neutrons not interacting at all and interacting neutrons with energy deposited in the iron,
%lost from the edges, or simply falling below threshold. 
The same level of performance was not reached in the present experiment with consequences discussed below (Sec.~\ref{sec:eff}).

\subsection{High-efficiency volume} %text from NIM article, rewrite
\label{sec:higheff}

The neutron recognition in the LAND detector for the case of large neutron multiplicity and with the setup of the present experiment 
has been investigated in detail with results reported in Ref.~\cite{piotr_nima}. As shown there, shadowing effects caused by 
the poles and yoke of the ALADIN magnet and the vacuum and detector chambers 
are visible in the hit distributions observed with LAND (cf. Fig.~6 in Ref.~\cite{piotr_nima}). 
The expected azimuthal symmetry around the incoming-beam direction 
is observed up to lateral distances of about 70 cm. A more rapid decrease 
observed at vertical distances beyond 80 cm above or below the plane containing the beam axis coincides with the
shadow of the poles of the ALADIN magnet and the magnet chamber as projected from the target position. The horizontal distribution extends
further out in the direction away from the deflected beam. There, the projected shadow of the magnet is close to
the end of the detector. Both distributions start to drop more rapidly at about 15 cm from the end of the detector.
These edge effects are most likely caused by a reduced detection efficiency for the coincident signal at the far 
side of the paddle for hits located near one of the ends of a paddle. 
 
With the aim to work with a homogeneous detection efficiency, a high-efficiency volume was defined within LAND
and only events whose primary hits are located within this volume were accepted for further analysis. 
It had the form of a cube extending over 
the full length of LAND with the lateral dimensions 
\begin{equation}
-150 < x < 0~\cm, -70 < y < 70~\cm
\end{equation}
in the beam-oriented coordinate frame (the $x$ direction points to the left, $y$ up, and the $z$ direction coincides with 
the original direction of the incoming beam). 
The corresponding angular acceptance in the laboratory, as viewed from the target, 
is $0^{\circ}$ to $-8.72^{\circ}$ in horizontal and $\pm 4.09^{\circ}$ in vertical directions at the entrance plane 
and $0^{\circ}$ to $-7.92^{\circ}$ and $\pm 3.72^{\circ}$ at the rear plane of the high-efficiency volume.

\subsection{Geant4 calculations} 
\label{sec:geant4}

In addition to shadowing, neutron scattering by the various materials of the setup is expected to produce a neutron 
background downstream of the spectrometer and possibly hitting LAND. To determine its potential strength within the high-efficiency volume, 
a detailed study was performed within the Geant4 framework~\cite{geant4}. 
The interaction of neutrons with the detector itself is not taken into account in these calculations; only 
the conditions under which neutrons emitted from the target can approach the detector are examined.
The model of the experimental 
setup consisted of replicas of the ALADIN magnet, the vacuum and detector chambers, and the time-of-flight wall, 
all with their accurate positions and dimensions but with a simplified geometry that omitted minor technical details.

\begin{figure}
\centerline{\includegraphics[width=7.0cm]{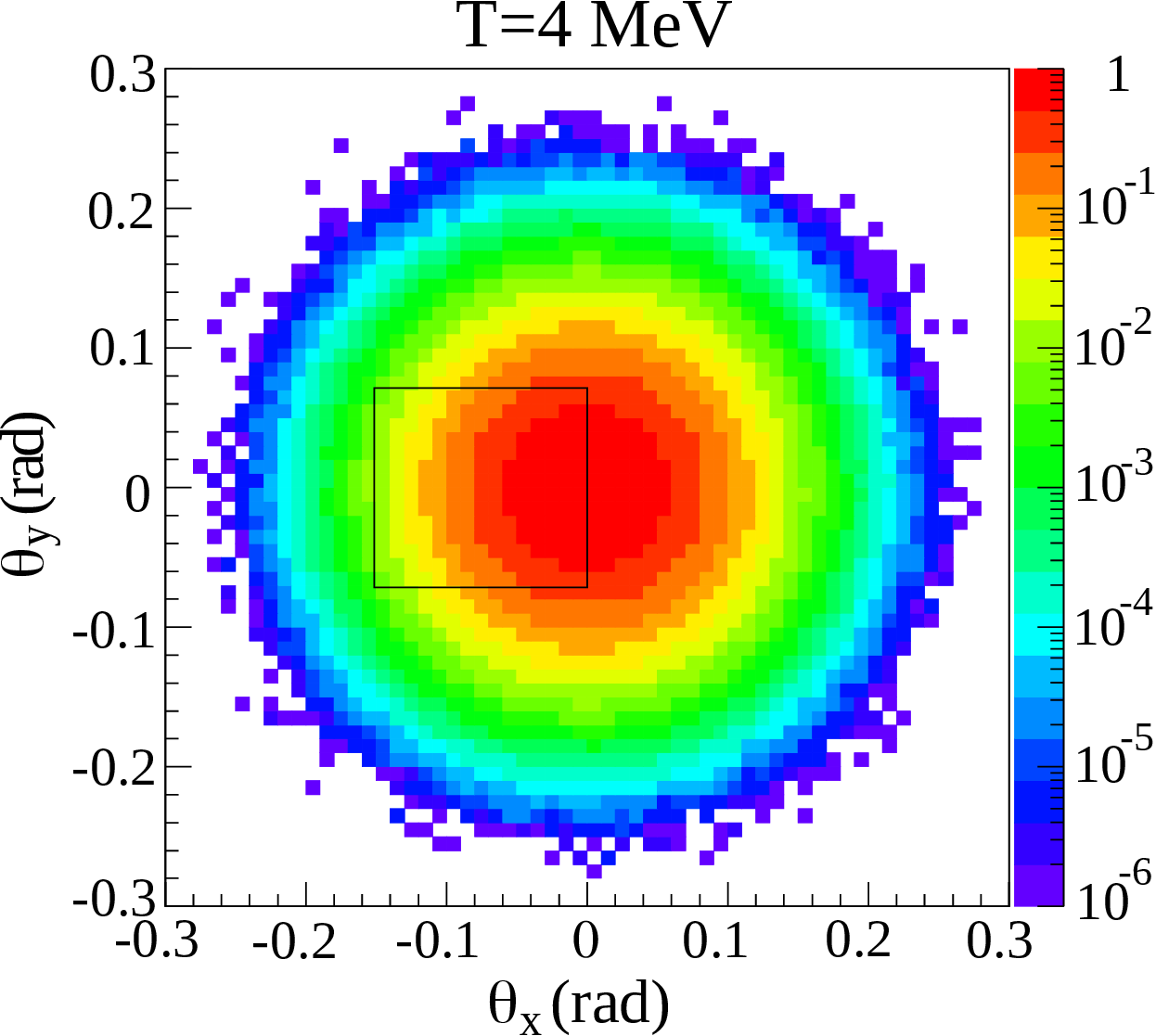}}
%\centerline{\includegraphics[width=6.5cm]{figures_wieloch/fig3.eps}}
\caption{\small{Probability distribution with respect to the beam direction of the neutron intensity of a thermal source
of temperature $T = 4$~MeV, located at the target position and  moving with the velocity of a projectile with 600 MeV/nucleon incident energy. 
The black square represents the high-efficiency area plotted in the chosen coordinate system in which
the negative $x$ direction viewed from the target position points to the right.}}
\label{fig:source4MeV}
\end{figure}     %fig. 3

The performed calculations fall into three classes. As a first step, pencil beams of neutrons with energies 
between 400 and 800 MeV were started from the target position, homogeneously distributed within a cone of 
polar angle $\theta_{\rm lab} \le 15^{\circ}$. 
Their trajectories and those of produced secondary particles were followed, if they were not absorbed earlier, 
until they passed through a test plane perpendicular to the original beam direction at the location of the front plane of LAND.

Within this plane, a high-efficiency area was defined with the lateral dimensions of the high-efficiency volume. 
From the calculated results, the probabilities for reaching the test plane and the high-efficiency area without any or 
after specific scattering processes were determined. The results in graphical form are given in the Appendix.

For the second type of calculations, thermal sources of neutrons moving with beam velocity were placed at the target position 
and their histories up to the test plane followed in the calculations. 
The chosen temperatures $T = 2-6$~MeV cover the interval of temperatures found for the identified spectator source of neutrons
(see Sec.~\ref{sec:thermal}).
The mere projection on the test plane 
of a source with temperature $T = 4$~MeV, i.e. as obtained with all material removed, is shown in Fig.~\ref{fig:source4MeV}. 
The black square representing the high-efficiency area covers roughly one half of the central part of the source, 
sufficiently large to determine the main source properties in that case. More precisely, 42.4\% of the particles of a source with 4 MeV
temperature are emitted in the direction of the high-efficiency area, a value decreasing from 45\% to 40\% for sources with temperatures 
increasing from 3 MeV to 5 MeV. 
% the latter being near the upper limit of source temperatures observed in this experiment.

\begin{figure}
\centerline{\includegraphics[width=6.5cm]{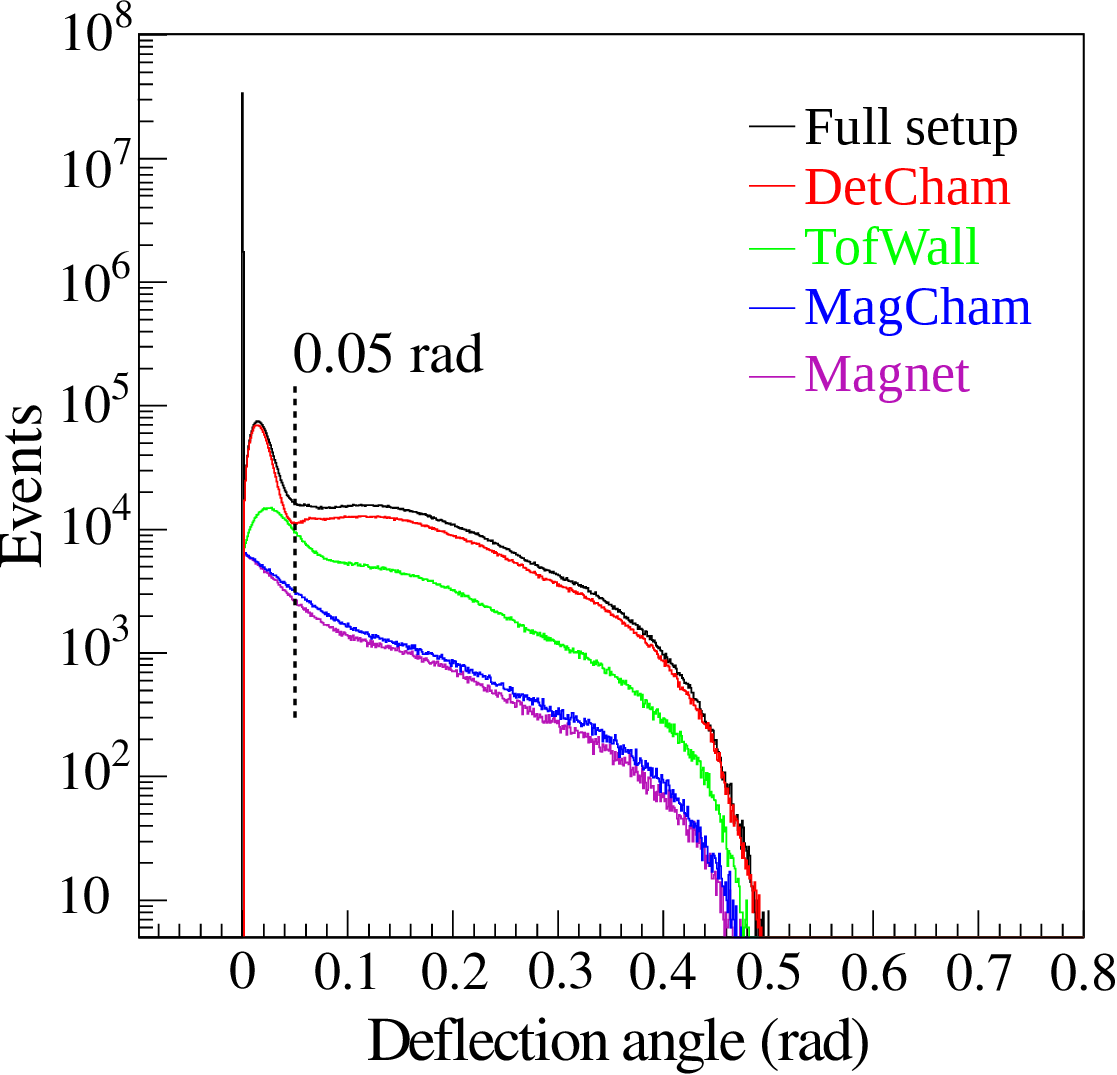}}
%\centerline{\includegraphics[width=6.5cm]{neutrons_fig4_new.eps}}
\caption{\small{Deflection angles of neutrons emitted in the direction of the high-efficiency area 
from a 4-MeV thermal source moving with beam velocity. 
The results are obtained from simulations for a source containing 10$^8$ neutrons in total with the full setup (black histogram) and with
individual parts of the setup as indicated, specifically the magnet (magenta), the magnet chamber (MagCham in blue), the time-of-flight
wall (TofWall in green), and the detector chamber (DetCham in red). The peak of non-interacting events, only shown for the full setup,
contains $3.56 \times 10^7$ events.
}}
\label{fig:scatter4MeV}
\end{figure}     %fig. 4

The probability of scattering processes suffered by neutrons of a 4-MeV source is illustrated in Fig.~\ref{fig:scatter4MeV}
by showing the distribution of deflection angles 
representing the apparent change of the emission angle when viewing the displacement at the test plane from the target position.
Deflection angles different from zero are caused by scattering processes 
experienced along the trajectories from the target to the test plane containing the front plane of LAND. 
For the figure, only the 42.4\% of neutrons emitted in the direction of the high-efficiency area are considered. 
We find that very few are absorbed on the way, and that altogether %$40.8 \cdot 10^6 / 42.4 \cdot 10^6$, i.e. 
96\% arrive at the test plane. 
A fraction of %$35.6 \cdot 10^6 / 40.8 \cdot 10^6$, i.e. 
87\% of them
reach the test plane in front of LAND unscattered as indicated by the peak at the origin in the distribution of deflection angles.
The small bump at deflections up to 0.05 rad corresponding to $\approx 3^{\circ}$ contains 
5\% of the neutrons arriving at the test plane, and only the remaining 8\% suffer interactions leading to larger deflections 
from their original directions. 
The most probable deflection corresponding to the maximum of the bump at $\approx 0.015$ rad is 15 cm at the test plane 9.78 m downstream.
This is small relative to the dimensions of the high-efficiency area, and more than 70\% of the scattered neutrons reach the
high-efficiency area according to the calculations for the 4-MeV source.

The locations at which the interactions occur are indicated as well in Fig.~\ref{fig:scatter4MeV}. The colored histograms show the distributions
of deflections obtained from calculations with virtual configurations containing only specific elements of the setup.
Very few interactions in the magnet or in the magnet chamber are experienced by neutrons emitted toward the high-efficiency area.
The contribution made by the time-of-flight wall is larger and exhibits the bump at small deflections. 
The main contribution originates from interactions in the exit flange of the detector chamber.

\begin{figure}
\centerline{\includegraphics[width=7.0cm]{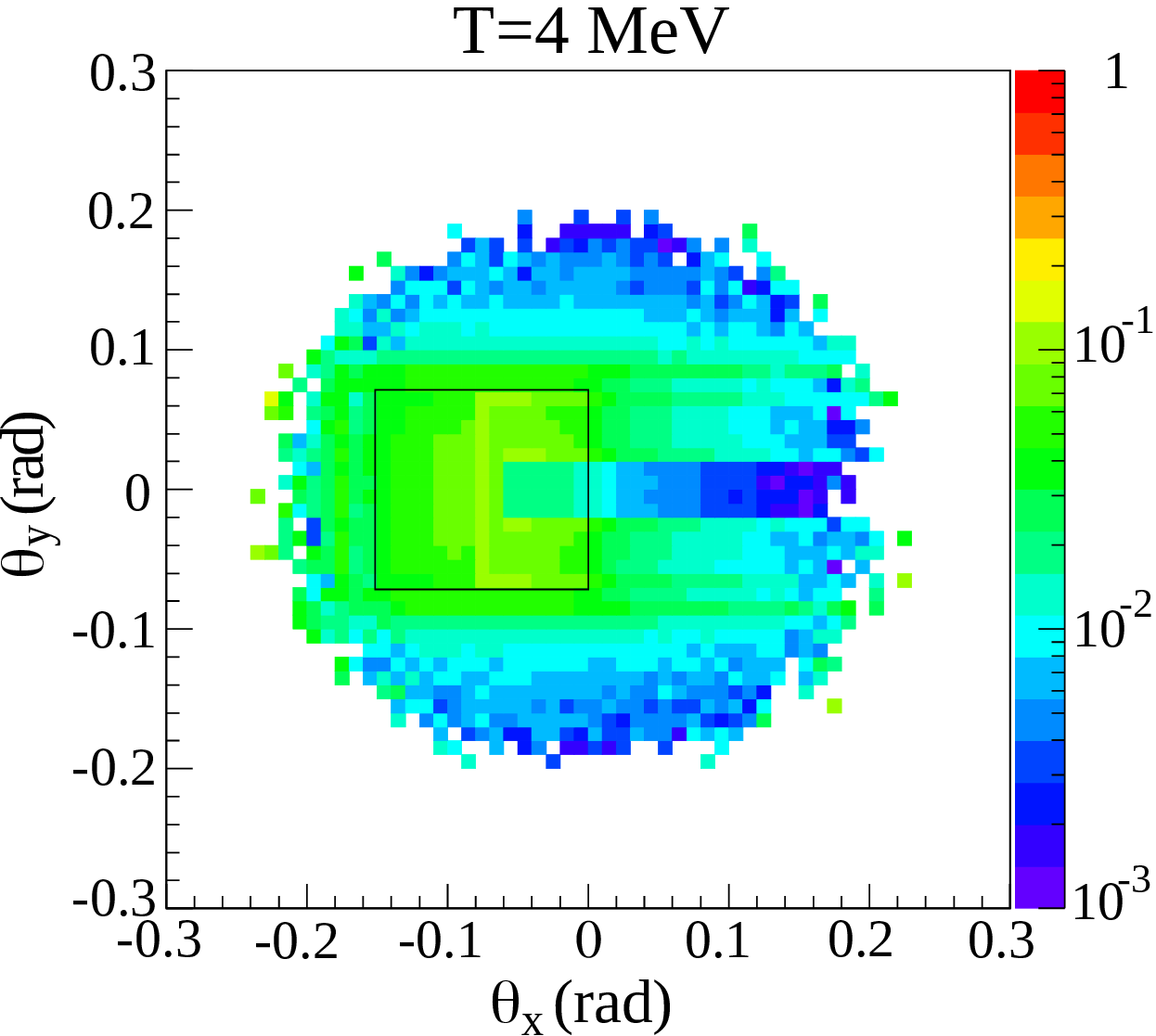}}
%\centerline{\includegraphics[width=6.5cm]{figures_wieloch/fig5.eps}}
\caption{\small{Origin of scattered neutrons reaching the high-efficiency volume of LAND: the color code represents the
probability for a neutron emitted with original direction ($\theta_x, \theta_y$) to enter the high-efficiency area of 
the test plane (black square) following a scattering process. The calculations are performed for a 4-MeV thermal source
moving with beam velocity whose emission pattern is shown in Fig.~\protect\ref{fig:source4MeV}.
}}
\label{fig:nogood4MeV}
\end{figure}     %fig. 5

The distribution of the original directions of all scattered neutrons that finally pass through the high-efficiency area of 
the test plane are shown in Fig.~\ref{fig:nogood4MeV}. The color code represents the probability of entering the high efficiency area
for a neutron starting with a given direction ($\theta_x, \theta_y$). 
Structure elements of the rear wall of the ALADIN detector chamber are
clearly recognized. The narrow horizontal band with reduced scatter probability is caused by the thin exit window for the beam. 
It is 1.98-m wide, 28-cm high, made from 1-mm steel, and welded at beam height into a thicker exit flange mounted on the rear wall 
(Fig.~\ref{fig:setup}). The rim structure of this flange is taken into account in the model of the experimental setup. 
It causes the vertical zone of higher scattering probability passing through the middle of the high-efficiency area 
at $\theta_x \approx -0.08$~rad.

Neutrons which, without scattering, 
would miss the high efficiency area have a probability of about 1--2\% to enter it, indicated by the light-blue color 
code outside the black square. Neutrons already starting toward the high-efficiency area have a probability of $\approx 10\%$ to scatter 
and to still end up there, indicated by the mostly light green color code within the black square. 
The majority of them belongs to the group of neutrons 
forming the prominent bump with deflection angles of up to 0.05 rad (Fig.~\ref{fig:scatter4MeV}). 
The integrated yield as shown in Fig.~\ref{fig:nogood4MeV} amounts to nearly 4\% of the source intensity %precisely 3.63%
which, added to the 36\% of neutrons reaching the high-efficiency area without interactions, yields $\approx 40\%$ of the total source intensity.
It is not far from the 42.4\% originally emitted in directions contained in the high-efficiency area.
However, the calculations also indicate that the scattered neutrons may lose energy in the process,  
forming a tail of lower intensity extending from the peak at beam rapidity to lower rapidities. 
Scattered neutrons are counted in the neutron multiplicities determined from the recorded total visible energy but 
may not be correctly placed in the rapidity and transverse-momentum distributions of the spectator source of neutrons. 

The calculated rates of absorption, scattering, and background production are functions of the chosen
source temperature but, for the relevant temperature interval, remain within 5--10\% which
indicates the general level of uncertainty caused by the interactions with the setup material. It may be summarized as a 
value of $90 \pm 10\%$ for the ratio of neutrons reaching the high-efficiency volume of LAND versus the number of neutrons emitted toward it.
Within this margin, the ALADIN-LAND setup appears to be well suited for the performed measurements.

A third type of calculations performed with nuclear beams had the aim of determining the level of a potential background 
of secondary neutrons produced by projectile fragments interacting with the material of the setup. The paths of selected types of fragments,
starting from the target position and deflected by the magnetic field, were followed through the setup until they traversed the time-of-flight wall 
and the exit wall of the detector chamber. It was found that
most of the secondary neutrons are emitted from a position near the end of the detector chamber and into narrow cones in directions close
to that of the exiting beam. Their transverse momenta are insufficient to enter the high-efficiency volume of LAND with large probability.
More details are given in the Appendix.

\subsection{Neutron identification and analysis}
\label{sec:nanal}

\begin{figure}
\centerline{\includegraphics[width=8cm]{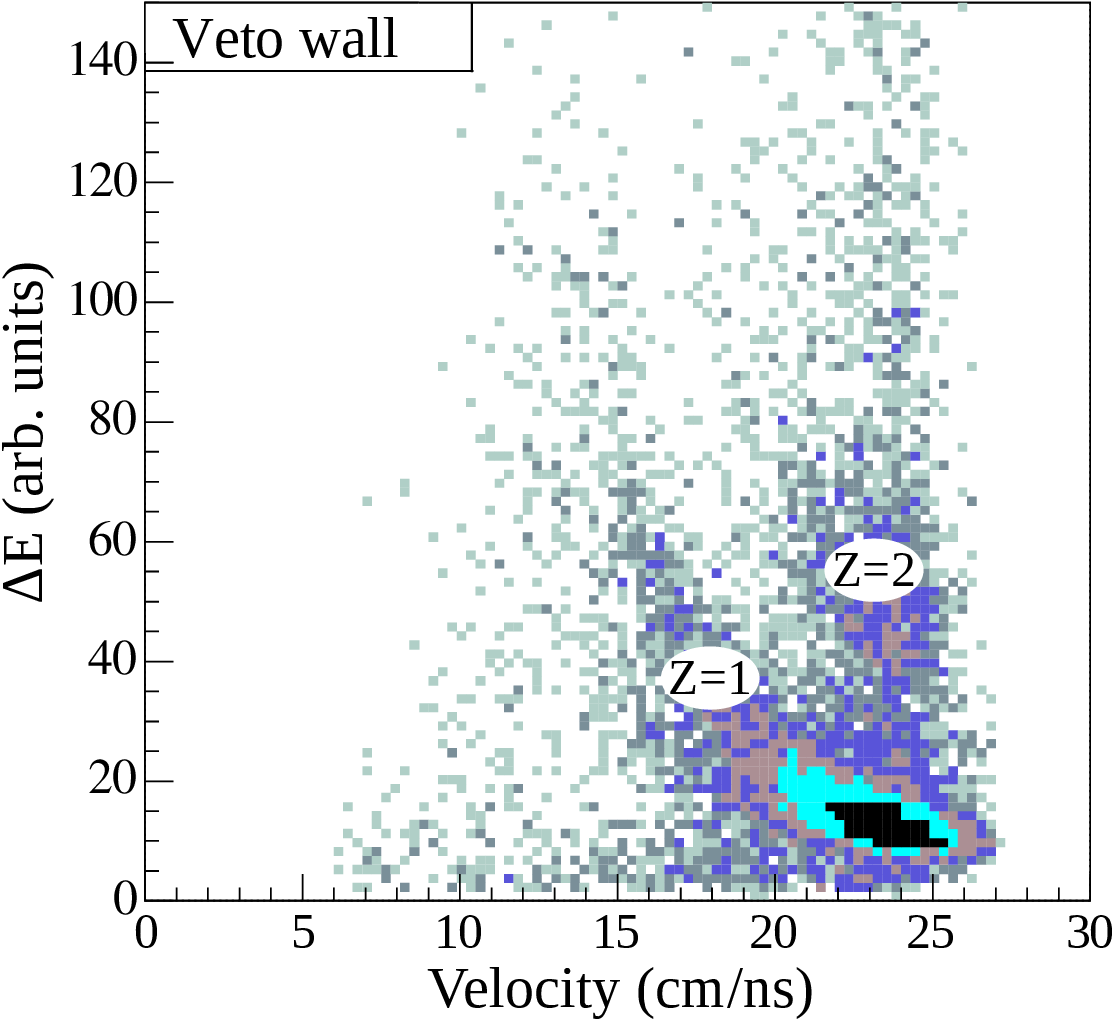}}
%\centerline{\includegraphics[width=8cm]{neutrons_fig6.eps}}
%\centerline{\includegraphics[width=8cm]{figures_wieloch/Evis_vs_Velocity_new_v4.eps}}
\caption{\small{Distribution of energy-loss signals $\Delta E$ measured with the veto wall in front of the LAND detector
as a function of the particle velocity. 
The registered charged particles are mainly protons with velocities close to the beam velocity of 23.8 cm/ns
and depositing approximately 1.4 MeV in the detector.}}
\label{fig:veto}
\end{figure}     %fig. 6

\begin{figure}
\centerline{\includegraphics[width=8.0cm]{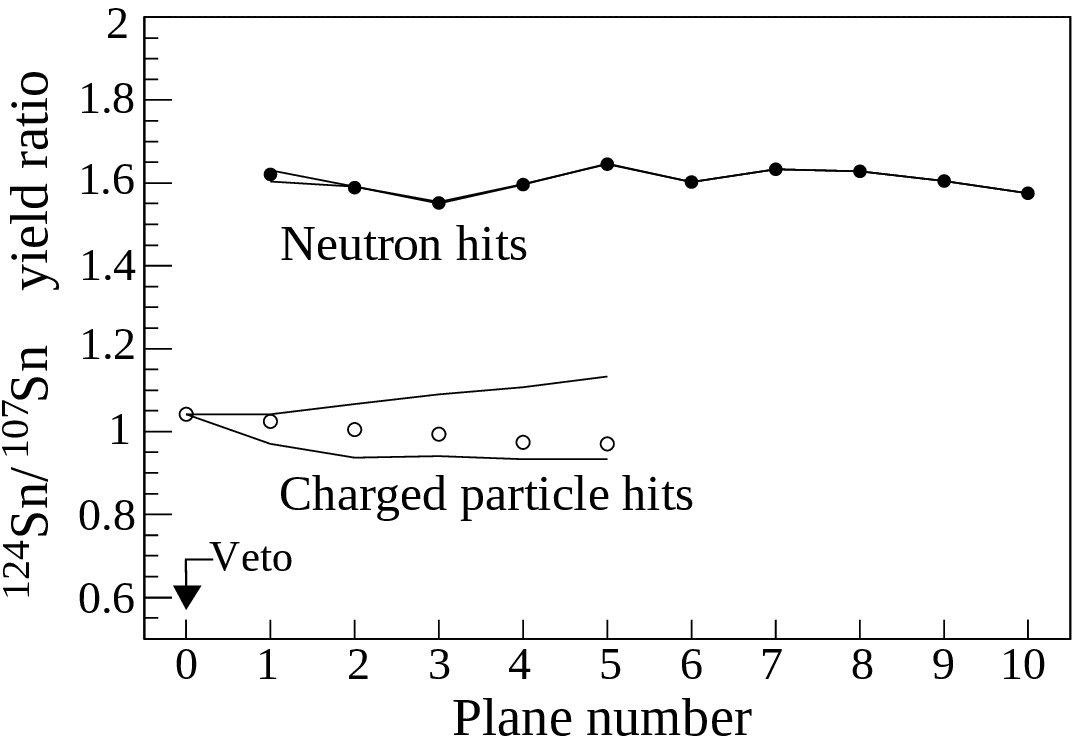}}
%\centerline{\includegraphics[width=8.0cm]{figures_wieloch/hits_vs_plane_new_v3.eps}}
\caption{\small{Ratios of registered hit multiplicities for neutrons (filled circles) and 
charged particles (open circles) from reactions with $^{124}$Sn and $^{107}$Sn projectiles 
as a function of the LAND plane number. Plane 0 represents the veto wall.
Solid lines correspond to upper and lower limits 
as obtained from the analysis (see text). They practically coincide for neutrons.}}
\label{fig:hits}
\end{figure}

A cluster-recognition algorithm is used to identify individual particles within the 
hit distributions registered with LAND~\cite{piotr_nima}. Charged particles are recognized by their
energy loss signals deposited in the veto wall (Fig.~\ref{fig:veto}). They consist of mainly hydrogen isotopes 
because heavier products emitted in forward direction into narrower angular cones
do no longer reach LAND after having passed the ALADIN magnetic field.
In the clustering procedures used for neutrons and charged particles, also the differences of their interaction 
with the LAND detector elements are taken into account. 
Neutrons produce secondary showers of ionizing particles in nuclear reactions with the detector 
material while charged particles, in addition, produce light signals by directly ionizing
the plastic layers along their flight paths. 
Charged-particle hits are strongly correlated in time and space, consecutively
following along the particle trajectory and possibly also ending in reaction showers similar to 
those generated by neutrons. 
According to their energy-loss in the detector material, 600-MeV protons do not penetrate
deeper into LAND than through the first five planes while hits from neutron events are 
distributed throughout the full detector volume (Fig.~\ref{fig:hits}).

     %fig. 7

The techniques used to analyze the LAND data are described in detail in Ref.~\cite{piotr_nima},
while here only the main analysis schemes are presented.
The cluster analysis begins by identifying charged-particle clusters by their correlation
with hits in the veto plane and by their consecutive appearance in the first few LAND planes.
The remaining hits are attributed to neutron events. In the clustering procedure for neutrons,
called the shower tracking algorithm (STA) in~\cite{piotr_nima},
hits are considered as correlated if they fall within a space and time window

\begin{eqnarray}
\label{eq:correlation}
-19 & < & \Delta x < 19~\cm, \nonumber \\
-19 & < & \Delta y < 19~\cm, \\
 -1 & < & \Delta t < 3~\ns    \nonumber
\end{eqnarray}

\noindent but are not required to belong to consecutive planes.
The search for clusters starts with the hits in the first plane and the search for a
correlated hit in one of the following planes and is continued with the iterative search for 
further correlated members of the cluster.
This procedure is then repeated by starting with remaining hits in the second LAND plane
and continued until all clusters are identified. 

The properties of a particle represented by a cluster of hits are determined by the 
position and time of the primary hit from which its velocity vector is calculated.
With the time resolution of $\Delta t \approx 650$~ps (FWHM),
including the time spread resulting from the depth of the paddles, and flight paths of 
about 10~m, the velocity resolution is $\Delta \beta/\beta \approx 1.6\%$.  
%new numbers; uncertainty of longitudinal position within a paddle increases time resolution from .250 to .278 ns (22.02.2023)

\subsection{Treatment of single hits}
\label{sec:single}

Not all hits are part of a cluster. When the clustering procedure is completed, a number of single hits may be left, isolated in space or time
so that no correlation with other hits was found. They are partly single-hit neutron events,
as expected on the basis of the shower statistics, but may also result from the following
technical inefficiencies:\\
(1) Missing paddles: A small number of paddles ($\approx 15\%$) were not properly functioning during the
experiment. It may have the effect that unobserved hits cause the separation of larger 
clusters into two or more parts, thereby generating isolated hits.\\
(2) Multiple hits: When two hits are registered in one paddle they are seen as
one hit with a false position. Also its time is usually altered so that existing correlations
may be destroyed. \\
(3) A neutron may react with the LAND material at two or more locations distant from each
other. These interactions will produce separate clusters or hits that are indistinguishable from
patterns generated by two or more neutrons. 

\begin{figure}
\centerline{\includegraphics[width=8.0cm]{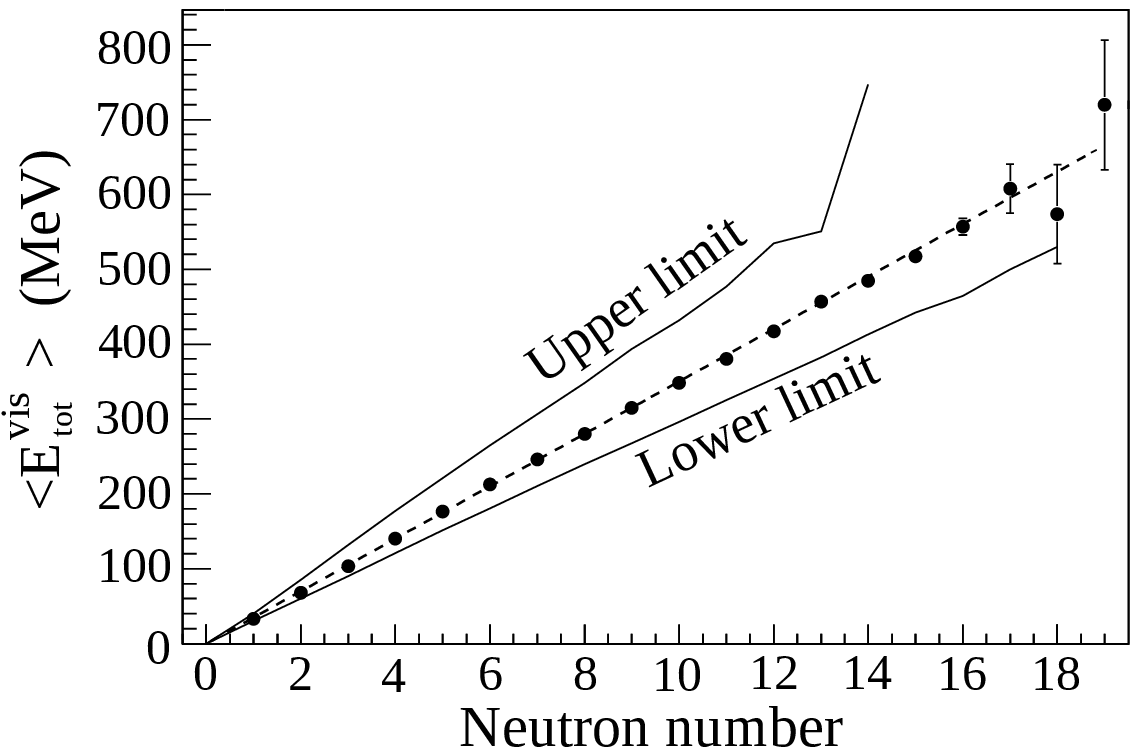}}
%\centerline{\includegraphics[width=8.0cm]{figures_wieloch/evis_vs_n_new_v3.eps}}
\caption{\small{The mean total visible energy $\langle E_{\rm tot}^{\rm vis} \rangle$ as a function of the neutron number obtained from
the AVE procedure (filled circles) for the case of $^{124}$Sn projectiles. 
Solid lines represent the upper and lower limits of $\langle E_{\rm tot}^{\rm vis} \rangle$ as obtained with the MIN and MAX procedures,
respectively. 
The dashed line corresponds to a mean value of $\langle E_{\rm tot}^{\rm vis} \rangle = 35$~MeV per 
neutron.}}
\label{fig:evis}
\end{figure}     %fig. 8

In order to assess the magnitudes of these effects, the data were analyzed with assumptions 
chosen to either maximize or minimize the number of resulting neutron events within 
reasonable limits (see Ref.~\cite{piotr_nima} for a detailed description). Counting all
single hits as neutron events produces an upper limit labeled ``MAX" in the following.
The number of single hits and thus of identified neutrons can be reduced by relaxing the requirements for a
correlation given in Eq.~(\ref{eq:correlation}). The results labeled ``MIN" in the following were produced by requiring only
the relaxed condition $-7 < \Delta t < 4$~ns in time and none of the conditions in space. 
Similarly, conditions in between these extremes were used to generate a most likely result
labeled ``AVE" (for average).

The effects of the different analysis procedures cancel for some observables. Examples are
the ratios of the multiplicities of hits assigned to neutron events as a function of the
plane number for the reactions of $^{124}$Sn and $^{107}$Sn projectiles (Fig.~\ref{fig:hits}).
The ratios of proton hits were found to depend on the choice of the procedure because effects
(1) and (2) may also disturb the recognition of charged particles. The independence of the plane number 
is best realized with the AVE conditions.

\begin{figure}
\centerline{\includegraphics[width=8.0cm]{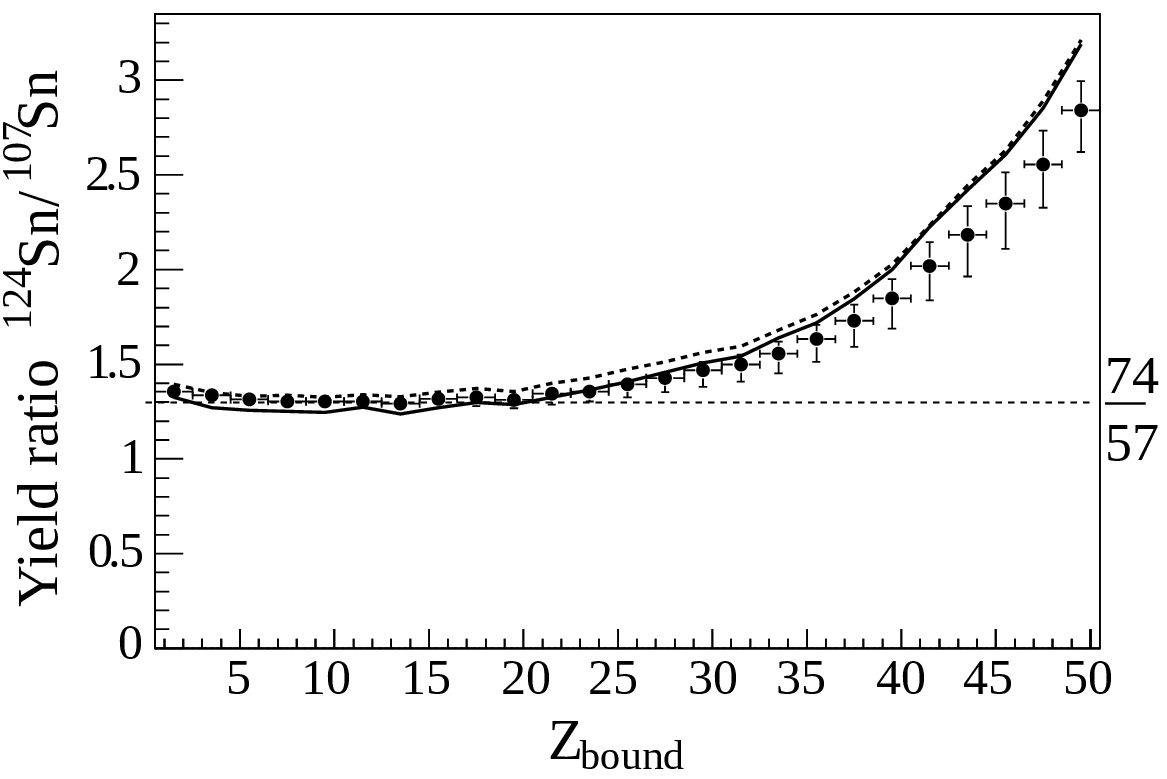}}
%\centerline{\includegraphics[width=8.5cm]{figures_wieloch/comp_new_v4.eps}}
\caption{\small{The ratios of the mean visible energies  $\langle E_{\rm tot}^{\rm vis} \rangle$ 
(solid line), of the hit numbers $N_{\rm hit}$ (dashed line) and of the neutron numbers $N$ 
from the clustering procedure (points) as obtained for reactions of the $^{124}$Sn and $^{107}$Sn 
projectiles as a function of $\zbound$. 
The error bars represent the upper and lower limits obtained with the MIN and MAX procedures.
The dashed horizontal line marks the ratio 74/57 = 1.30 of the neutron numbers of the two Sn projectiles.
}}
\label{fig:comp}
\end{figure}     %fig. 9

The magnitude of the differences between the procedures is illustrated in Fig.~\ref{fig:evis}.
The total visible energy $E_{\rm tot}^{\rm vis}$ obtained by integrating the recorded
pulse heights should, on average, rise in proportion to the number of detected
neutrons. The expected linearity is observed for each of the three procedures but the mean
visible energy per neutron is considerably smaller for the MAX than for the MIN procedures.
A mean value $\langle E_{\rm tot}^{\rm vis} \rangle = 35$~MeV is obtained with the AVE procedure.
An absolute calibration for this value at the time of the experiment does not exist for LAND. 
However, in the previous
experiment S107, tagged neutrons from the breakup of deuteron beams at various energies were used to
study the response of the detector to single neutrons~\cite{boretzky03,gsi1992}. With this data set, multi-neutron events 
were generated with the technique of event mixing and, subsequently, analyzed with the present
clustering procedures~\cite{piotr_nima}. The most satisfactory results were obtained with 
the AVE procedure, which was therefore adopted as the most realistic choice.

As a further test, we show in Fig.~\ref{fig:comp} the ratios of the measured mean visible energies 
$\langle E_{\rm tot}^{\rm vis} \rangle$, of the hit numbers $N_{\rm hit}$, and of the 
neutron numbers $N$ from the three clustering procedures as obtained from the two 
reactions with $^{124}$Sn and $^{107}$Sn projectiles as a function of $\zbound$. 
In the limit of small $\zbound$, i.e., large excitation energies, this ratio approaches the
value 74/57 = 1.30, i.e. the ratio of the neutron numbers of the two projectiles, 
%This is expected for the vaporization limit, corresponding to the disintegration of the projectiles
%into mainly nucleons and light charged particles. 
and the ratios of all three observables coincide. On the opposite side,
for large $\zbound$, the evaporation from excited projectile residues is the dominant source
of neutrons. Considerably more neutrons are emitted from the neutron-rich $\sn$ than from
the neutron-poor $\snrad$ residues. It explains why the $\zbound$ averaged ratio 1.6 shown in 
Fig.~\ref{fig:hits} is larger than 74/57. For large $\zbound$, the ratio of the neutron 
numbers determined with the clustering procedure is lower by up to 10\% than those of the 
global observables $\langle E_{\rm tot}^{\rm vis} \rangle$ and $N_{\rm hit}$. 
%This difference may be considered as representative for the systematic error associated with the chosen cluster-finding method.

\subsection{LAND efficiency}
\label{sec:eff}

A calibration of the LAND efficiency for neutrons at the time of the experiment does not exist.
As the data analysis indicates, the performance of the detector had been slightly degraded. 
About 15\% of the paddles were not functioning properly, mainly in the rear part of the detector.
More importantly, the thresholds for registering hits had increased.
The analysis of the measured hit distribution for one-neutron events in comparison with the S107 results obtained with 
the fully functional detector led to the conclusion that the hit-registration probability had decreased to approximately 
one half of its original value~\cite{piotr_nima}. The corresponding detection 
efficiency for one-neutron events was estimated as only 73\%.

The obtained value, however, did not take into account that the breaking
of clusters by unobserved 
hits into two or more separate parts, mostly single hits, 
may increase the neutron multiplicity returned by the applied clustering algorithm. In that respect, it may be considered a lower limit,
similar to the MIN choice in STA discussed in Sec.~\ref{sec:single} and represented by the upper limit displayed in Fig.~\ref{fig:evis}. 
The alternative AVE choice, yielding multiplicities larger by a factor $\approx 1.25$ than the MIN option, was found to best account for
the actual inefficiencies of LAND and to lead to reliable multiplicity estimates in simulations using actual spatial and temporal hit 
distributions measured in the present experiment (cf. Fig.~3 in~\cite{piotr_nima}). 
The resulting most probable detection efficiency at the time of the experiment thus amounts to $0.73 \times 1.25 \approx 0.9$.

There is even a tendency, in particular for the larger multiplicities 5 to 15 encountered here, for a slight 
overestimation of the order of one up to two neutrons, corresponding to 10\% to 20\% in this range of multiplicities. 
The effect persists in simulations modeling properties of LAND at the time of the experiment (cf. Fig.~5 in~\cite{piotr_nima}). 
By its magnitude, it may even partly compensate for the losses 
caused by the finite LAND efficiency of 0.9. 
We therefore adopt the results 
obtained with the AVE conditions without further corrections 
as the most realistic choice for the present experiment. 

The uncertainty of this procedure is clearly significant.
Even if the MIN and MAX procedures are rejected as extreme options, it may still be estimated to be of the order of $\pm 15\%$. 
Together with the probability of $90 \pm 10\%$ for neutrons reaching the high-efficiency volume of LAND
(Sec.~\ref{sec:geant4}), we arrive at a value of $0.9 \pm 0.2$ for the overall detection efficiency applicable to this experiment.
As a common factor, it only mildly affects relative results obtained in comparing the studied reactions (cf. Fig.~\ref{fig:comp}).

\section{EXPERIMENTAL RESULTS}
\label{sec:res}

\subsection{Reaction characteristics}
\label{sec:react}

\begin{figure} [htb]
\centerline{\includegraphics[width=8cm]{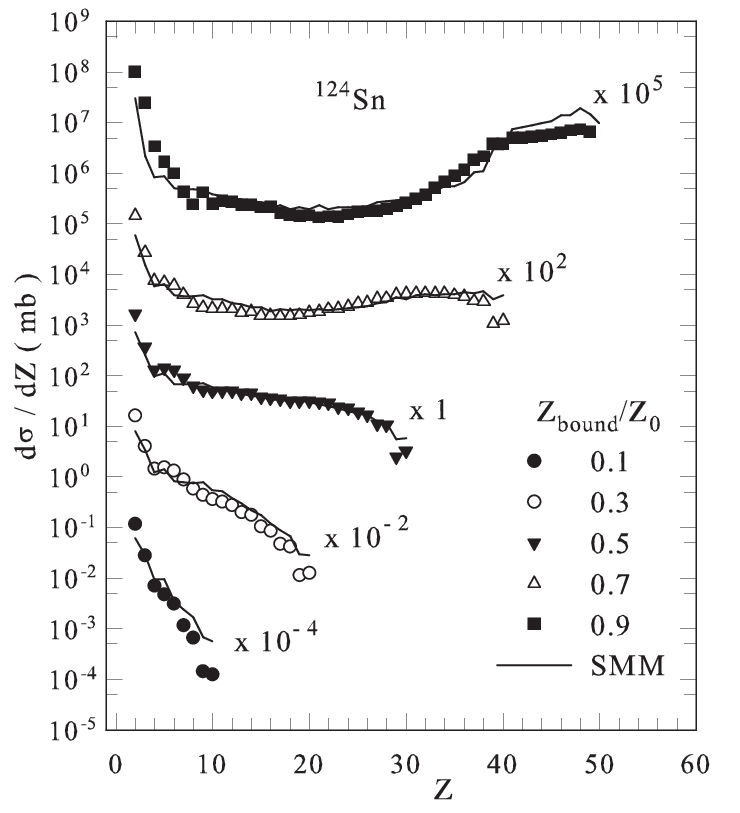}}
\caption{\small{Experimental cross sections d$\sigma$/d$Z$ for the fragment production following
collisions of $^{124}$Sn projectiles with a natural Sn target at 600 MeV/nucleon, sorted into five intervals of $Z_{\rm bound}/Z_0$ with centers
as indicated and width 0.2 (symbols) in comparison with normalized SMM calculations (lines). 
$Z_0 = 50$ represents the atomic number of the $^{124}$Sn projectiles.
The scale factors used for displaying the cross sections are indicated 
(reprinted with permission from Ref.~\protect\cite{ogul11}; Copyright $\copyright$ 2011 by the American Physical Society).
}}
\label{fig:z124}
\end{figure} %fig. 10

As an introduction, two figures will be used to briefly characterize the studied reactions. The fragmentation 
of the projectile spectators after the initial stage of the collision was investigated in detail in Ref.~\cite{ogul11}. 
As an example taken from this work, 
the experimental cross sections d$\sigma$/d$Z$ for the fragment production following collisions of $^{124}$Sn projectiles 
are shown in Fig.~\ref{fig:z124}, sorted into five intervals of $Z_{\rm bound}$. The charge distributions 
evolve from a so-called 'U-shaped' distribution at large impact parameters, with the production of heavy residues and light fragments 
in asymmetric binary decays, through a power-law shaped 
to a rapidly dropping exponential spectrum in the bin of smallest impact parameter. 
This evolution is a well-known and characteristic phenomenon and expected as a manifestation of the nuclear liquid-gas phase
transition~\cite{smm,ogilvie91,gross90,kreutz,hauger00}.  
The thin lines in the figure represent the results of SMM calculations as described in Ref.~\cite{ogul11}. 
A comparable reproduction of 
the charge correlations characterizing the projectile fragmentation at the present energy was achieved with the 
dynamical isospin quantum molecular dynamics (IQMD) transport model~\cite{junsu18}. 

\begin{figure}     
%\centerline{\includegraphics[width=8.5cm]{figures_hakan/bertini.eps}}
\centerline{\includegraphics[width=8cm]{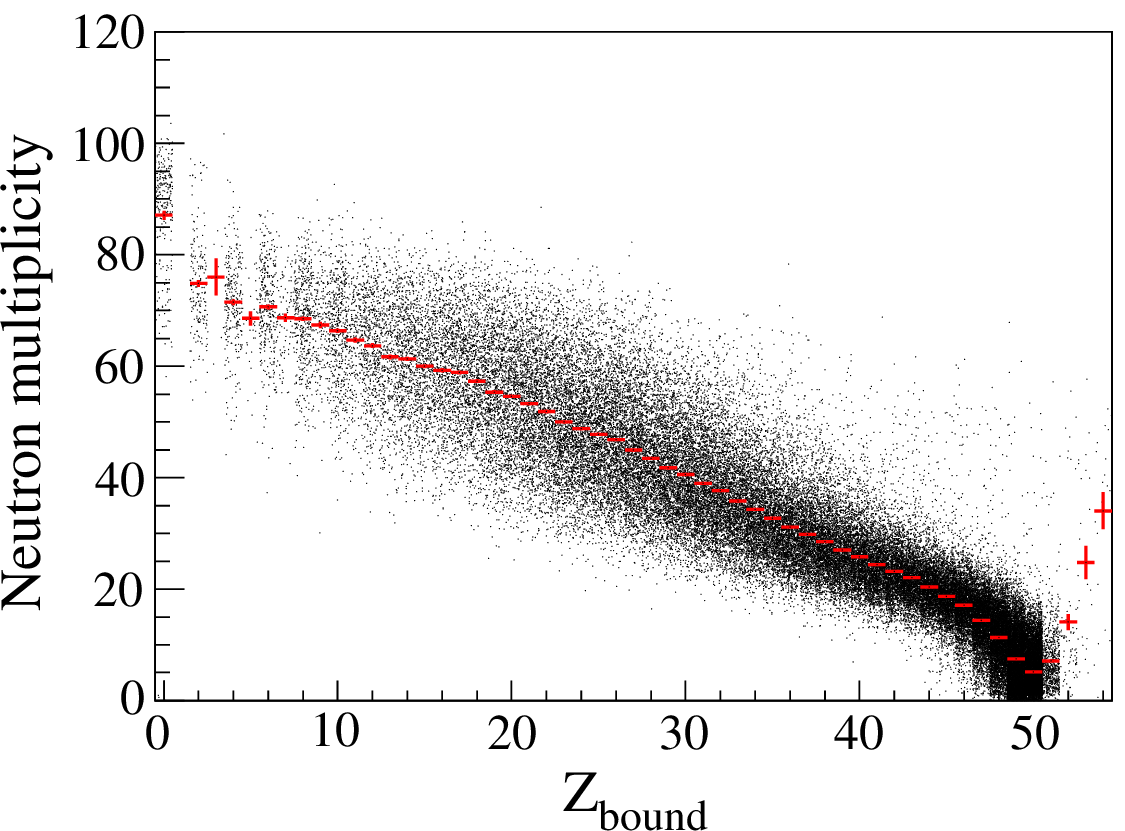}}
%\centerline{\includegraphics[width=8cm]{figures_wieloch/fig11_new.eps}}
\caption{\small{Neutron multiplicity as a function of $Z_{\rm bound}$ according to the Bertini cascade implemented in Geant4 
for the reaction $^{124}$Sn + Sn at 600 MeV/nucleon. A condition $E_{\rm lab} > 100$~MeV for neutrons to be counted is imposed. 
The short-dashed line (red) follows the mean values of the so obtained multiplicity for given $Z_{\rm bound}$.
}}
\label{fig:bertini}
\end{figure}     %fig. 11

The neutron multiplicities expected for these processes are shown in Fig.~\ref{fig:bertini} as a function of $Z_{\rm bound}$. 
The predictions were obtained with the Bertini cascade model as implemented in the Geant4 toolkit~\cite{bertini}. 
Besides the initial intranuclear cascade with excitons, the model treats preequilibrium emissions, 
Fermi breakup for excited nuclei with $A < 12$ and $Z < 6$, and
fission and evaporation from the produced spectator nuclei. The calculations were performed by simply directing a beam of $^{124}$Sn projectiles 
on a Sn target of thickness 0.5 mm within the Geant framework without taking any properties of the setup into account. 
Only neutrons with kinetic energies exceeding $E_{\rm lab} = 100$~MeV were counted, 
a condition meant to exclude neutrons from secondary deexcitations of target fragments. The odd-even structure at small $Z_{\rm bound}$
follows from the fact that $Z_{\rm bound}$, by definition, cannot be equal to 1 and that events containing a single light fragment of odd $Z$
without accompanying $\alpha$~particles are apparently rare. 
The odd-even structure is also visible in the measured reaction cross sections~\cite{ogul11}.

The calculated reaction cross section $\sigma_{\rm react} = 4.52$~b compares well with the value measured in the experiment~\cite{ogul11}. 
In peripheral collisions, the calculated neutron multiplicities reach rapidly values near 20 from where they increase smoothly 
with decreasing $Z_{\rm bound}$ to values close to 80 for the number of free neutrons in the most central collisions. 
This latter number is consistent with expectations based on the measured composition of the light-particle source in central collisions. 
For the $^{197}$Au + $^{197}$Au reaction at 400 MeV/nucleon, the relative yields of p:d:t:$^3$He:$^4$He were found to be 1:0.65:0.4:0.2:0.2, 
as measured by the FOPI Collaboration~\cite{reisdorf10}. Neglecting heavier fragments and assuming the 
same composition for the $^{124}$Sn + Sn system, with a very similar $N/Z$ but at the higher energy 600 MeV/nucleon, 
leads to a composition containing 35 free protons, 65 protons and 72 neutrons bound in light charged particles, and 76 free neutrons.
However, very few of these fireball neutrons are likely to be emitted into the small solid angle in forward direction covered by LAND.

\begin{figure} [tbh]
\centerline{\includegraphics[width=8.5cm]{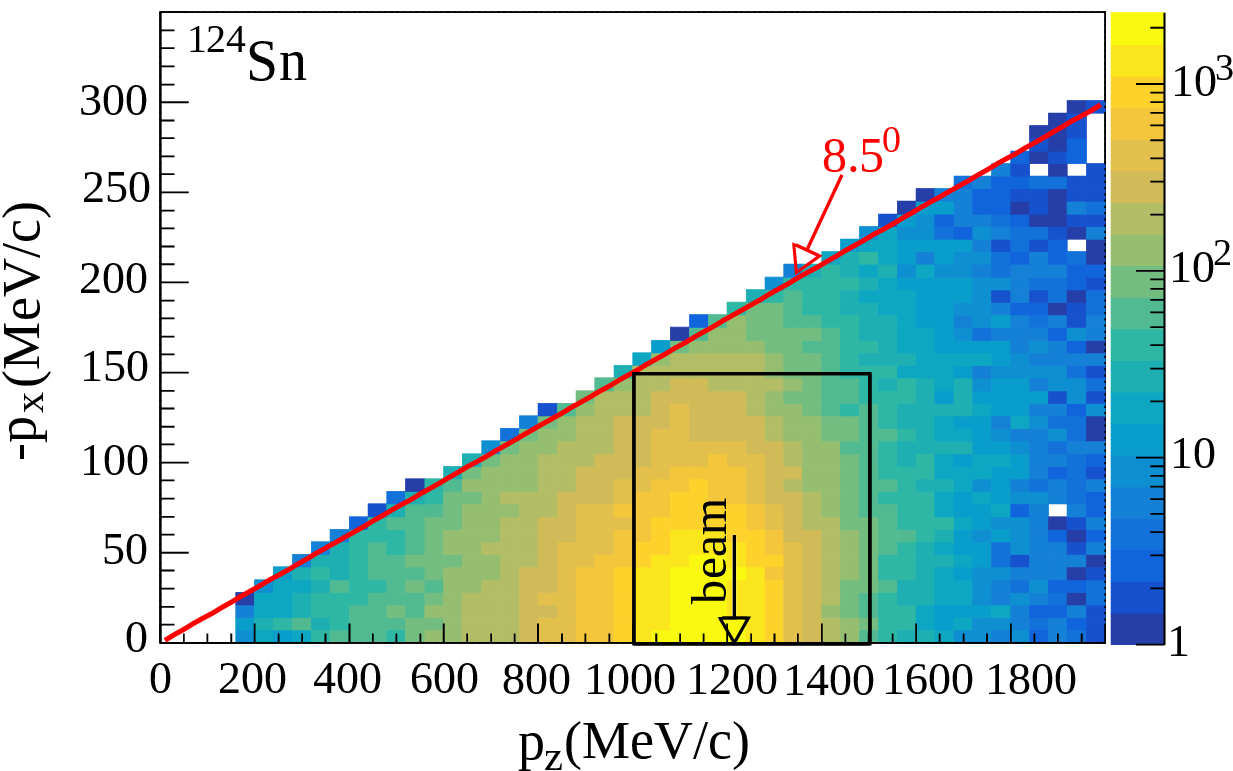}}
%\centerline{\includegraphics[width=8.5cm]{figures_wieloch/source2_new_v4.eps}}
\caption{\small{Distribution of identified neutrons in the plane of transverse momentum $-p_x$
versus longitudinal momentum $p_z$ for the fragmentation of $^{124}$Sn projectiles
in collisions with natural Sn targets at 600 MeV/nucleon incident energy. The full
line (red) indicates the acceptance cut in the analysis. The black rectangle indicates the
two-dimensional interval $|p_x| \le 150$~MeV/$c$ and $1000 \le p_z \le 1500$~MeV/$c$ used
for the determination of the source temperature and multiplicity. 
}}
\label{fig:source}
\end{figure} %fig. 12

As will be shown below, the multiplicity of the spectator source of neutrons identified with LAND reaches its maximum for 
$Z_{\rm bound} \approx 35$ with values around 11 for the $^{124}$Sn projectile whereas it is around $30 - 35$ 
according to the Bertini model (Fig.~\ref{fig:bertini}). It is clear that, at the corresponding intermediate impact parameters, 
the intensity of the fireball is still significant but the number of free neutrons may also be somewhat overestimated with
the Bertini model, possibly by up to $10 - 15$ neutrons. In this range of $\zbound$, the multiplicity of intermediate
mass fragments is still around one or larger~\cite{ogul11}, and the observed charge spectrum stretches up to the evaporation 
regime around $Z = 30$ (Fig.~\ref{fig:z124}). This type of multifragmentation process with neutrons bound in heavier fragments
is not explicitly accounted for in the model. The calculations, nevertheless, indicate that the  
total number of free neutrons emitted in mid-peripheral collisions with $Z_{\rm bound} \ge 35$ and particularly also at peripheral
collisions with $Z_{\rm bound}$ up to nearly 50 is larger than 
that of the spectator neutron source experimentally observed at forward angles.

\subsection{Spectator source of neutrons}
\label{sec:source}

Longitudinal momenta were determined from the neutron time-of-flight measured with a
resolution of $\Delta t \approx 650$~ps (FWHM), corresponding to $\Delta p_z \approx 50$~MeV/$c$ or about 4\% for beam velocity neutrons. 
The accepted range of perpendicular momentum is largest in the horizontal direction but only on the side opposite to 
the deflected beam (cf. Figs.~\ref{fig:setup} and \ref{fig:source4MeV}), 
and is shown for $^{124}$Sn projectiles in Fig.~\ref{fig:source}. 
The spectator source is clearly visible and is centered at momenta slightly below the projectile 
momentum $p_{\rm proj} = 1216$~MeV/$c$ per nucleon. 
The interval in the plane of momenta $p_x$ vs $p_z$ 
used for determining the magnitude and apparent temperature of the neutron source is indicated. 
For the transverse dimension, an upper limit $-p_x \le 150$~MeV/$c$ was chosen which keeps the analysis interval within the
high-efficiency volume for longitudinal momenta $p_z \ge 1000$ ~MeV/$c$.
The rapid drop of the source intensity 
in the interval $50 \le |p_x| \le 100$~MeV/$c$ confirms that also in the vertical dimension the dominant part of the spectator 
neutron source is accepted within the high-efficiency volume with $|p_y| \le 87$~MeV/$c$ for beam-velocity neutrons.

\begin{figure} [tbh]
\centerline{\includegraphics[width=7.5cm]{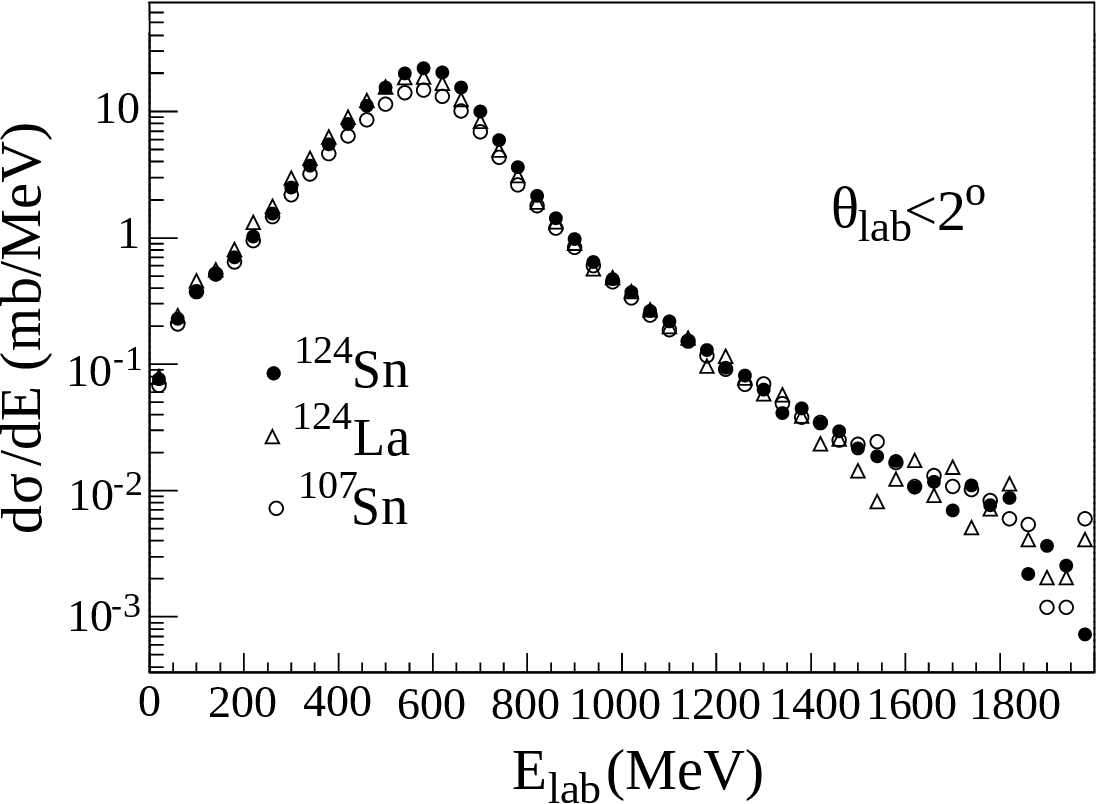}}
%\centerline{\includegraphics[width=8.0cm]{figures_wieloch/e0_n_new_points_new_v3.eps}}
\caption{\small{Inclusive differential cross section of neutrons in the laboratory detected within the solid angle $\theta_{\rm lab} < 2^{\circ}$
for the three projectiles as indicated. No corrections are applied to the recorded data (cf. Sec.~\protect\ref{sec:eff}).
}}
\label{fig:2deg}
\end{figure} %fig. 13

Absolute cross sections were obtained with the normalization determined for this experiment~\cite{ogul11}.
Differential production cross sections measured for neutrons within a cone $\theta_{\rm lab} < 2^{\circ}$ at very forward angles and 
integrated over the range of $\zbound \ge 2$ up to the projectile $Z$ are shown in Fig.~\ref{fig:2deg}. 
The choice of $\theta_{\rm lab} < 2^{\circ}$, much narrower than the acceptance of LAND in this experiment, was made to get close
to the geometry of the Berkeley experiments
of Madey {\it et al.}~\cite{madey81,madey85,baldwin92}, and to permit the comparison with their results (see Sec.~\ref{sec:bevalac}).

Here and in the following, no corrections are applied to the recorded data because precise correction factors were not obtained. 
According to the results presented in Sec.~\ref{sec:geant4}, the possible losses of neutrons between the target and the detector 
are of the order of $10 \pm 10\%$, somewhat depending on the angular distribution of the emitted neutrons. 
The LAND efficiency, on the other hand, is close to unity but is somewhat dependent on the event multiplicity (Sec.~\ref{sec:eff}).
According to the deduced overall detection efficiency $0.9 \pm 0.2$, the recorded cross sections and mean multiplicities
are possibly too low by about 10\% but are affected with uncertainties up to $\pm 20\%$. 

The spectra reach their maxima at laboratory energies 560 to 600 MeV and drop off rapidly toward higher and lower energies. 
In a thermal interpretation, their widths indicate temperatures of 4.5 to 5 MeV for the emitting sources, 
not taking into account a possible broadening caused by a finite distribution of the longitudinal velocities of the spectator sources. 
For reasons discussed below, the HWHM on the high-energy side was used for this estimate. The double differential 
cross section reaches a peak value d$\sigma/d\Omega$/dE = $5.5 \times 10^3$ mb/sr/MeV for $^{124}$Sn fragmentations and values smaller 
by about 30\% and 40\% for the cases of $^{124}$La and $^{107}$Sn, respectively. These differences are, again, larger than 
the relative differences of the neutron numbers of the three projectiles (cf. Figs.~\ref{fig:hits} and \ref{fig:comp}).

\subsection{Thermal properties}
\label{sec:thermal}

The momentum distributions and energy spectra shown in Figs.~\ref{fig:source} and~\ref{fig:2deg} 
suggest that the neutron emission from the spectator source has thermal
characteristics with Maxwellian distributions in the projectile frame. 
This is expected for neutrons evaporated from heavy projectile residues and should mainly reflect the internal temperature 
of the emitting fragment.
The effects of recoil and of the motion of the emitting source in the projectile rest frame are very small in this case.
The momentum transfer to the projectile residues is not discussed in Ref.~\cite{ogul11} but existing preliminary analyses indicate that it is
very similar to those reported for Au and Pb fragmentations in Refs.~\cite{schuett96,chance01,huentrup01}.
It adds to the momenta of neutrons emitted from lighter fragments. In the simplest approximation, its contribution to the observed neutron
temperature may be estimated as $T_f/A_f$ with $T_f$ being the temperature describing the fragment motion in the projectile rest frame and 
$A_f$ the mass number of the emitting fragment. With $T_f \approx 15$~MeV as deduced in Ref.~\cite{schuett96}, this contribution is indeed 
very small for emissions from heavy residues but may reach values of the order of MeV for emissions from lighter fragments. 
The resulting temperatures should thus be considered as effective temperatures of the projectile neutron source.

Because of their superior momentum resolution, the transverse components $p_{x}$ and $p_{y}$ were chosen for the following analysis.
The neutron density distribution in the $p_{y}$ vs $p_{x}$ plane
is assumed to be given by

\begin{equation}
\frac{\partial^{2}N}{\partial p_{x}\partial p_{y}}=\Ns\Gt(p_{x})\Gt(p_{y})+\Nb\beta(p_{x},p_{y})
\label{eq:density_background}
\end{equation}

\noindent where $\Ns$ is the measured neutron multiplicity, and $\Gt$ is a classical Maxwellian function of the form

\begin{equation}
\Gt(p_{i})=\frac{1}{\sqrt{2\pi mT}}\exp\left(-\frac{p_{i}^{2}}{2mT}\right),\, i=x,y
\label{eq:maxwell}
\end{equation}

\noindent with neutron mass $m$ and temperature parameter $T$. A potentially existing neutron background is taken into account 
with the added background term in Eq.~(\ref{eq:density_background}). Here
$\Nb$ is the total number of background neutrons and $\beta(p_{x},p_{y})$
is a background distribution function, normalized to unity. This function
is not known, however, and has been assumed 
to be approximately constant over the region dominated by the  emission from the projectile spectators.
The goal is to determine the three parameters temperature
$T$, source neutron multiplicity $\Ns$, and the background
neutron multiplicity $\Nb$.

To find the temperature parameter $T$, 
the distribution of one of the particle
momentum components is fitted with a Gaussian function added to a constant
background pedestal:

\begin{equation}
f(p_{x})=C_{1}\exp\left(-\frac{p_{x}^{2}}{2\sigma^{2}}\right)+C_{2}\label{fit}
\end{equation}
\label{eq:tempfit}

\noindent 
where $C_{1}$, $C_{2}$ and $\sigma$ are the fitted parameters. The comparison
with Eq.~(\ref{eq:maxwell}) gives a correspondence between the $\sigma$
parameter and the temperature: $\sigma=\sqrt{mT}$. 

\begin{figure}
\centerline{\includegraphics[width=7.5cm]{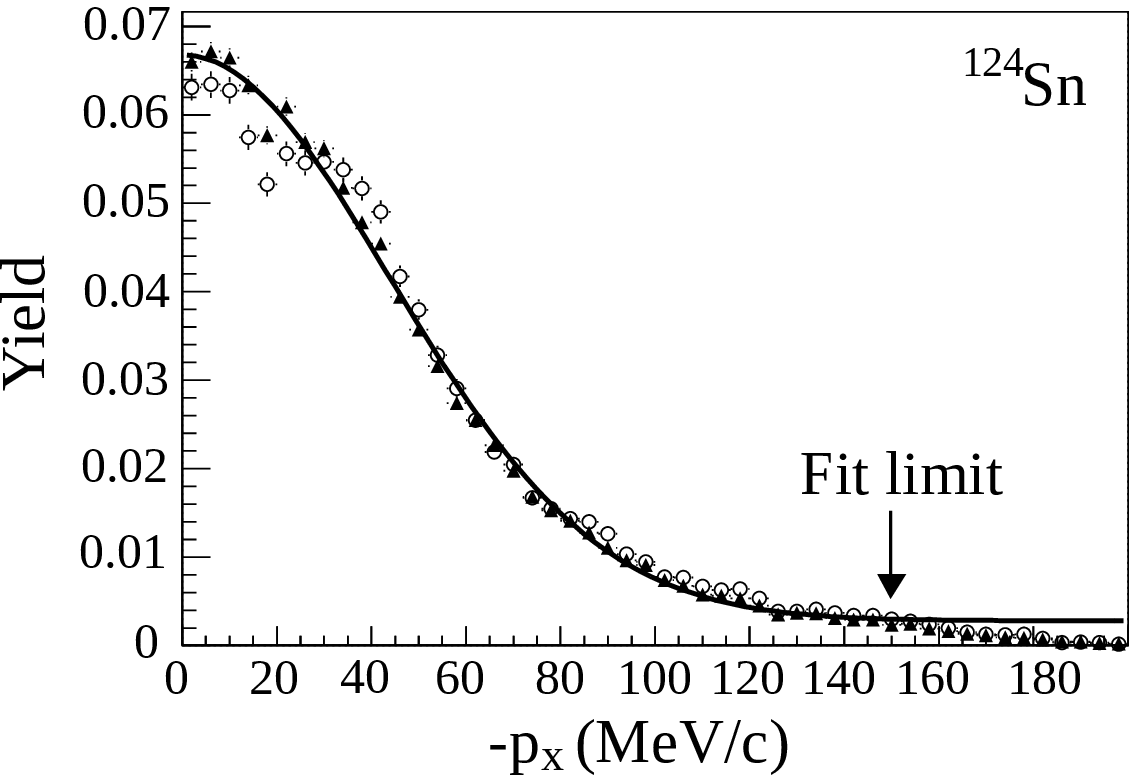}}
%\centerline{\includegraphics[width=8.0cm]{px_distribution.eps}}
\caption{\small{Distribution of transverse momentum $-p_x$ for neutrons from the fragmentation 
of $^{124}$Sn projectiles, selected with the condition $Z_{\rm bound} \ge 45$.
The open circles and the filled triangles represent the normalized yields of primary hits within a neutron 
cluster and of all hits, respectively.
The result of a fit with a Gaussian distribution corresponding to $T=2.04$~MeV superimposed on a 
constant background is represented by the solid line. The fit limit at $p_x = -150$~MeV/$c$ is indicated.
}}
\label{fig:pxfit}
\end{figure}    %fig. 14
% figure taken from Bormio 2011 in PoS

The transverse-momentum spectra for $|p_x| \le 150$~MeV/$c$ are indeed well described by Gaussian fits superimposed on a constant pedestal 
that may represent neutrons from a source with wider momentum distributions together with a background of scattered neutrons. 
An example of the results obtained for the fragmentation of $^{124}$Sn projectiles is shown in Fig.~\ref{fig:pxfit}. 
In the chosen rather peripheral case with $Z_{\rm bound} \ge 45$, neutrons can be assumed to be mainly emitted 
from excited residues (cf. Fig.~\ref{fig:z124}). The background level is evidently very low,
consistent with the expectation that scattering processes may cause the flight times to be longer
and to partly fall outside the chosen interval corresponding to $p_z > 1000$~MeV/$c$.
The fit limit $-p_x \le 150$~MeV/$c$ is equal to the upper limit of the analysis interval (Fig.~\ref{fig:source}).

\begin{figure}
\centerline{\includegraphics[width=7.5cm]{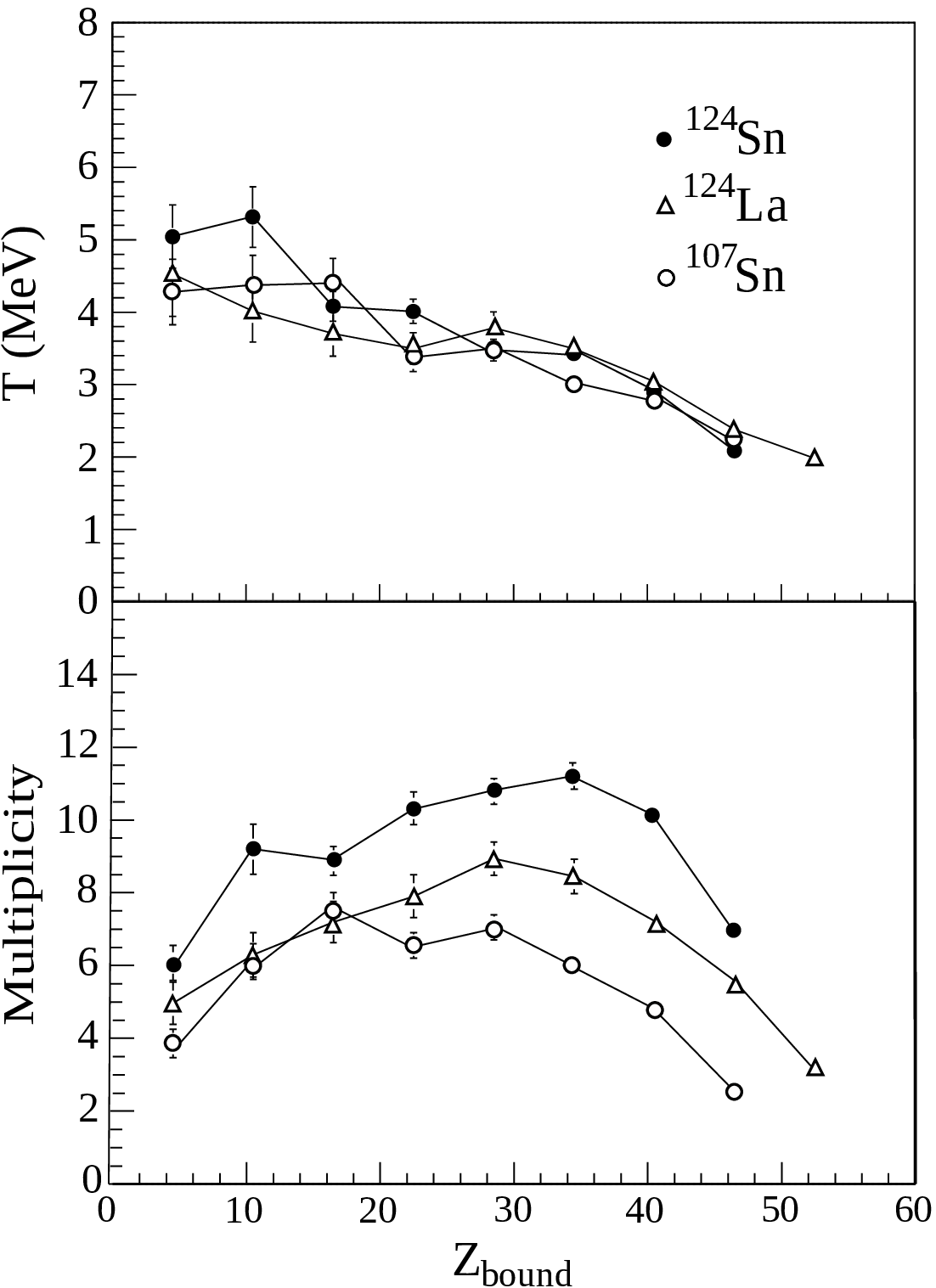}}
%\centerline{\includegraphics[width=8.0cm]{figures_wieloch/mult_temp_v4.eps}}
\caption{\small{Top: temperature parameter of the projectile source as a function
of $\zbound$ for $\sn$ (filled circles), $\la$ (triangles), and $\snrad$ (open circles) projectiles. 
Bottom: multiplicity of neutrons emitted from the projectile source as a function
of $\zbound$ for the same three cases. The vertical bars in both panels include the errors returned by the fitting routine
but not the overall uncertainty of the obtained multiplicities.
No corrections are applied to the recorded data (cf. Sec.~\protect\ref{sec:eff}).
}}
\label{fig:temp_mult}
\end{figure}    %fig. 15

\begin{figure} [tbh]
\centerline{\includegraphics[width=7.5cm]{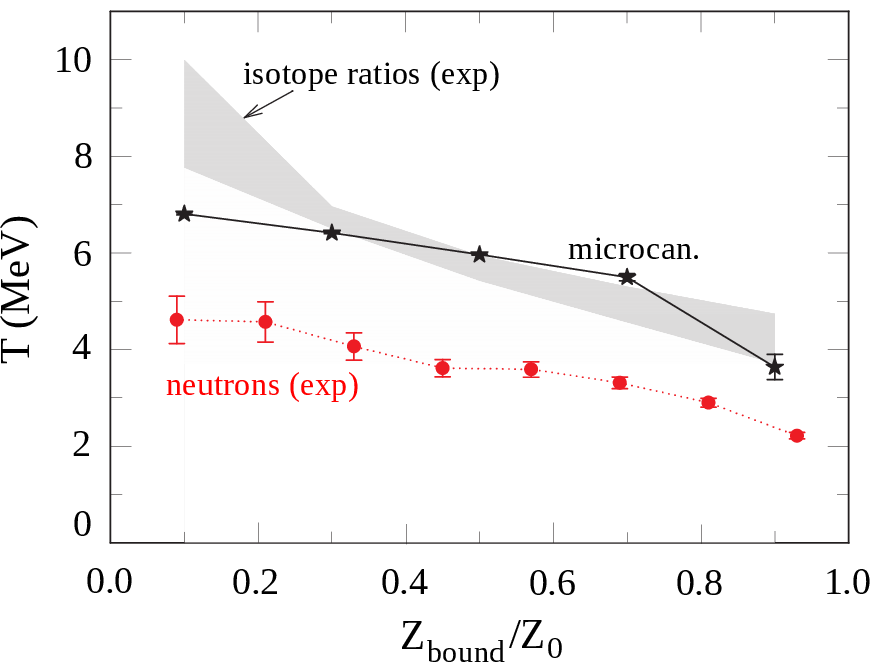}}
%\centerline{\includegraphics[width=8.0cm]{figures_smm/Figure19_final.eps}}
\caption{\small{Average temperature of the neutron sources of the three systems, determined from the measured momentum distributions, 
as a function of $\zbound$, normalized with respect to the projectile atomic number $Z_0$, (filled circles in red) 
in comparison with the mean microcanonical temperature of the SMM description, calculated with a symmetry coefficient 
$\gamma = 14$~MeV and averaged over the three projectile ensembles (stars in black). 
The average individual errors for the three systems are shown where they are larger than the symbols. 
The lines are meant to guide the eye.
The shaded band represents the experimental isotope temperatures reported in Ref.~\protect\cite{sfienti09}, 
including the spread of the six values of the $T_{\rm HeLi}$ and $T_{\rm BeLi}$ temperatures for the three systems 
(figure adapted from Ref.~\protect\cite{ogul11}). 
}}
\label{fig:smmtemp}
\end{figure}

The temperatures determined with this method are shown in Fig.~\ref{fig:temp_mult} (top) as a function of $\zbound$.  
They are found to be quite similar for all three reactions, smoothly increasing from about 2 MeV at large $\zbound$ to values 
between 4 and 5 MeV for the most central collisions recorded in this experiment, there with uncertainties of up to 0.5 MeV. 
Part of this rise may be related to the source motion and recoil effects discussed above.
Over the full range of $\zbound$, the neutron
temperatures are lower by typically about 2 MeV than the chemical breakup temperatures of 4 to 8 MeV determined for the present 
reactions~\cite{sfienti09}. For $\zbound = 20 - 30$, the interval for which the multiplicity of intermediate-mass fragments 
reaches its maximum (cf. Fig.~\ref{fig:z124}), the neutron temperatures are between $T = 3.5$ and 4.0 MeV while the chemical breakup temperature 
is $T \approx 6$~MeV (Fig.~\ref{fig:smmtemp}). 

 %fig. 16

Lower temperatures are expected as average values in the case of deexcitation cascades that
lead to the ground states of the emitting fragmentation products. However, a similar reduction has not been observed for 
the evaporation component identified in proton spectra measured for $^{197}$Au fragmentations at 1000 MeV/nucleon~\cite{odeh00}. 
The proton temperature $T_p \approx 6$~MeV in the range of maximum fragment production is similar to the double-isotope 
temperatures determined for these reactions~\cite{poch95,xi97,traut07}. 
For $\alpha$ particles from the same reaction, an evaporation component with $T_{\alpha} = 5 \pm 2$~MeV
has been identified in the measured kinetic energy spectra~\cite{traut07}. In spite of the partly large errors, 
the observed temperatures of evaporation-type light particles 
seem compatible with an interpretation identifying them as properties of secondary emissions with neutrons dominating in 
the later stages.

\subsection{Integrated neutron yields}
\label{sec:integ}

By integrating over the three-dimensional isotropic Gaussian source with the width parameter determined from the $p_x$ spectra,
the multiplicities, i.e. the number of neutrons emitted from the projectile-spectator source, were obtained.
These neutron multiplicities as a function of $\zbound$ are shown in the bottom panel of Fig.~\ref{fig:temp_mult}. 
They exhibit broad maxima in the $\zbound$ interval of approximately 20 to 40. For larger $\zbound$ 
the excitation energies appear to be lower, so that the neutron emission drops accordingly. For smaller $\zbound$,
the excitation energy per nucleon may be larger but the source size becomes smaller as a result of the more central impact 
parameter. At small $\zbound$, the neutron multiplicities correspond to the $N/Z$ ratio of the projectile. 
They are, on average, 30\% higher for the case of $^{124}$Sn than for the two neutron-poor projectiles, 
a value close to their difference in $N/Z$ as already indicated in Fig.~\ref{fig:comp}. 
In more peripheral collisions, for $\zbound > 25$, the neutron multiplicities exhibit a strong dependence
on the neutron content of the incoming projectiles, stronger than in proportion to their total neutron numbers
which are equal to effectively 57.7, 67.5, and to 74 for the three cases.
However, after normalization with respect to the mass number $A_0$ of the projectile and plotted as a function of $\zbound /Z_0$, 
the multiplicities of the two neutron poor systems describe a common curve (cf. Ref.~\cite{sfienti05}).

\begin{figure} [tbh]
\centerline{\includegraphics[width=7.5cm]{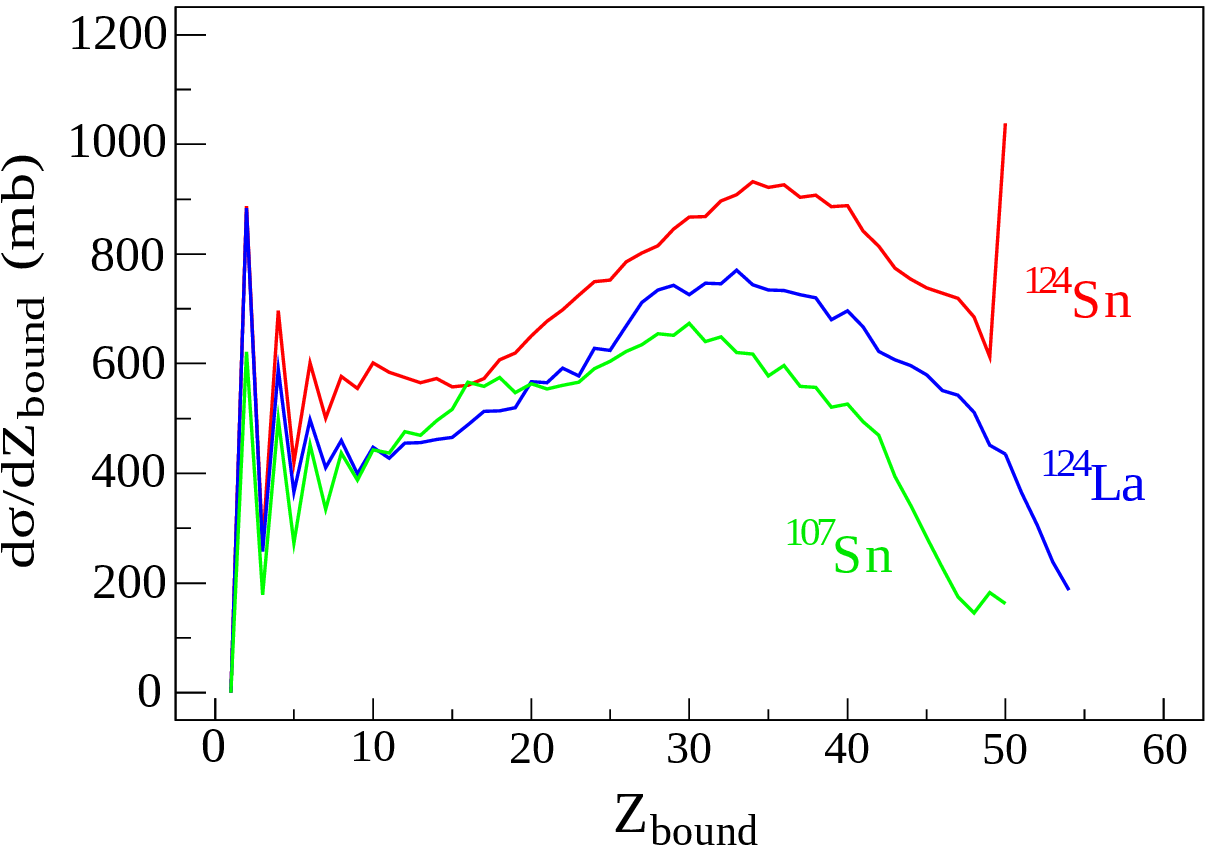}}
%\centerline{\includegraphics[width=8.0cm]{z_bound_mb_neut.eps}}
\caption{\small{Production cross sections of the identified spectator source of neutrons as a function of $\zbound$ for
the three projectiles $^{124}$Sn, $^{124}$La, and $^{107}$Sn (see text for the uncertainties).
}}
\label{fig:xsect}
\end{figure} %fig. 17

Production cross sections of the identified spectator neutron source were obtained by multiplying the multiplicities with
the measured event cross sections shown in Fig. 4 of Ref.~\cite{ogul11}. For that purpose, the multiplicities were
linearly interpolated, and slightly extrapolated at both ends, to obtain a smooth variation with $\zbound$. 
The results are shown in Fig.~\ref{fig:xsect}. 
At small $\zbound$, 
the cross sections exhibit the alternating structure caused by the definition of $\zbound$ as explained above and in 
Ref.~\cite{ogul11}. With increasing $\zbound$, they essentially reflect the differences of the multiplicities. 
The $\zbound$-integrated cross sections are large and amount to 35 b for $^{124}$Sn and
30 and 24 b for the neutron poor $^{124}$La and $^{107}$Sn, respectively. These values are affected with the overall uncertainty
of the measured neutron multiplicities (Sec.~\ref{sec:eff}) and by the inefficiencies of the experimental trigger at large $\zbound$
discussed in Ref.~\cite{ogul11}.

\section{STATISTICAL ANALYSIS}
\label{sec:sta}

%where ? Model studies confirm that these $\langle A/Z \rangle$ values are also representative for the 
%spectator systems emerging after the initial stages of the reaction~\cite{LeFevre05,Botvina02}.

%check what has been left out from Nihal's intro + sections
%compare description with Pshenichnov paper arXiv:0911.2017 in Sagwan refreport.

\subsection{Ensemble calculations} %from Nihal
\label{sec:ensemble}

The statistical interpretation of the fragmentation processes observed in the present reactions was performed within the 
statistical multifragmentation model (SMM, Ref.~\cite{smm}) and reported in Ref.~\cite{ogul11}.  
The statistical breakup and decay were calculated for parametrized ensembles of excited sources representing the variety of 
excited spectator nuclei expected in a participant-spectator scenario. Following the scheme developed in Ref.~\cite{botvina95}, 
the parameters of the ensemble were determined empirically by searching for an optimum reproduction of the measured fragment charge 
distributions and correlations.
As an example, the experimental charge distributions,
measured for $^{124}$Sn and shown in Fig.~\ref{fig:z124}, are well reproduced. Equally good descriptions have been 
achieved for the cases of $^{107}$Sn and $^{124}$La projectiles~\cite{ogul11}.

The microcanonical temperatures obtained for the disintegrating sources were found to coincide with the double-isotope-yield
temperatures deduced from He, Li, and Be fragment yields~\cite{sfienti09} as shown in Fig.~\ref{fig:smmtemp}.
Effects of the parameters of the liquid-drop description of produced fragments were investigated as well. 
In particular, the coefficient $\gamma$ of the symmetry term had to be modified, so as to provide an adequate description 
of the measured isotopic distributions of fragments with $Z \le 10$.

\subsection{Neutron multiplicities and temperatures}
\label{sec:nmult}

\begin{figure} [tbh]
\centerline{\includegraphics[width=7.5cm]{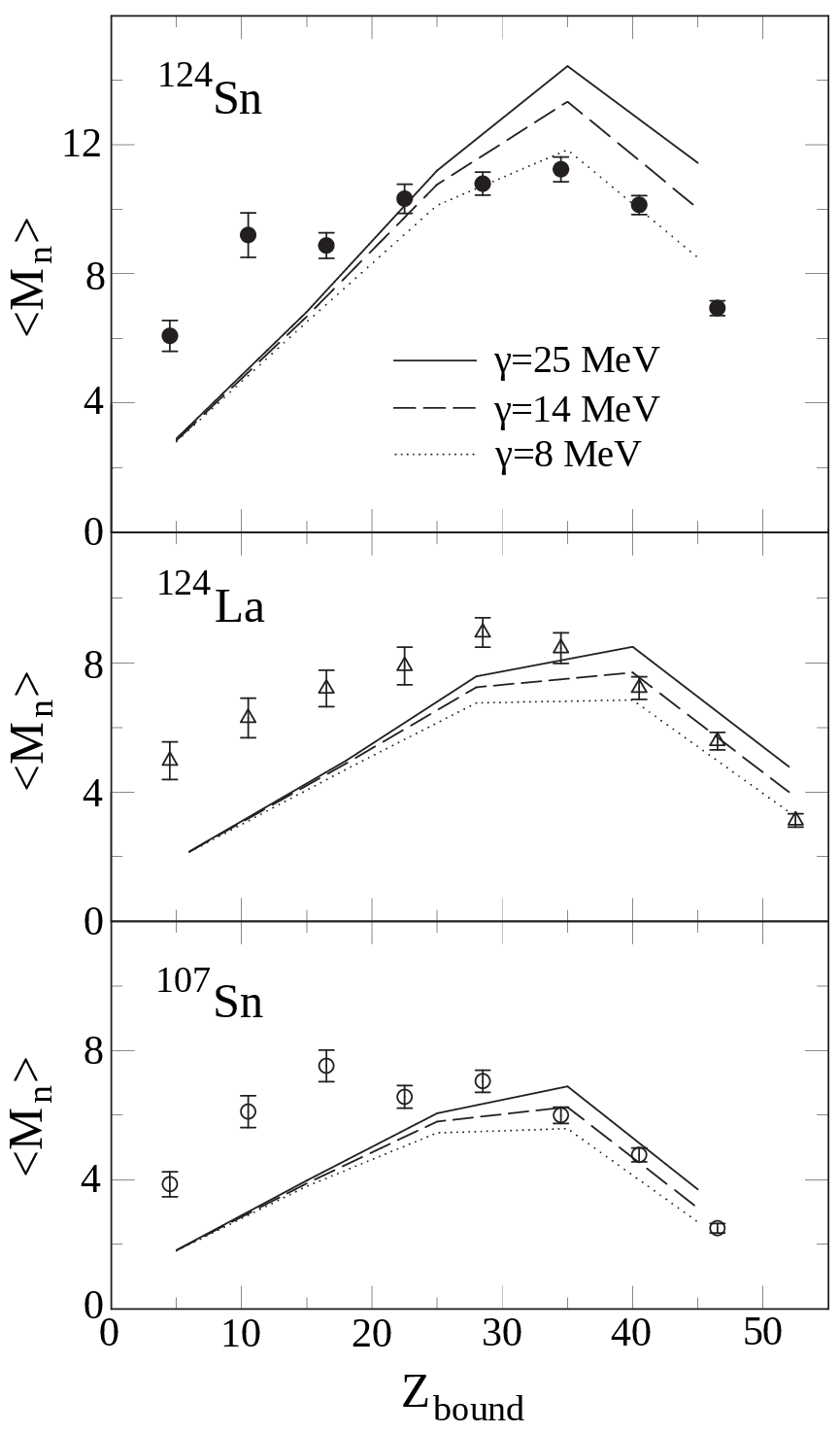}}
%\centerline{\includegraphics[width=7.5cm]{figures_smm/Fig16_errorbars_yscale_equal.eps}}
\caption{\small{Mean neutron multiplicity $\langle M_n \rangle$  of the spectator sources
produced in the fragmentation of $^{124}$Sn (top panel), $^{124}$La 
(middle panel), and $^{107}$Sn (bottom panel) as a function  of $Z_{\rm bound}$. 
The experimental data, presented in Fig.~\ref{fig:temp_mult}, are given by the symbols. The results of the SMM ensemble 
calculations, shown for values of the symmetry energy coefficient $\gamma$ = 25, 
14, and 8 MeV, are represented by the full, dashed, and dotted lines as indicated. 
The displayed experimental errors do not include the overall uncertainty discussed in Sec.~\ref{sec:eff}.
}}
\label{fig:smmnmult}
\end{figure} %fig. 18

The SMM predictions for the mean neutron multiplicity as a function of $Z_{\rm bound}$ are shown in Fig.~\ref{fig:smmnmult}. 
They exhibit the observed global behavior with maxima appearing at intermediate $\zbound$ but are below the experimental values 
at small $\zbound$. 
At large $\zbound$, in the regime from U-shaped to power-law spectra (Fig.~\ref{fig:z124}), 
the experimental multiplicities are reproduced rather well. In addition, a significant 
sensitivity to the coefficient $\gamma$ chosen for the symmetry term in the liquid-drop description of the produced fragments is observed, 
most prominently for the case of the neutron-rich $^{124}$Sn. 
%If the correction of $\approx 10\%$ expected from the Geant4 model study 
%is taken into account, the calculations with the reduced value $\gamma = 14~\MeV$ may come closest to the measured multiplicities. 
The comparison is consistent with the observation that a reduced value for $\gamma$ is required for successfully
reproducing the yields of neutron rich fragments~\cite{ogul11}. A more precise determination of $\gamma$ is precluded
by the overall uncertainty $\pm 20\%$ of the experimental neutron multiplicities (Sec.~\ref{sec:eff}).

\begin{figure} [tbh]
\centerline{\includegraphics[width=7.5cm]{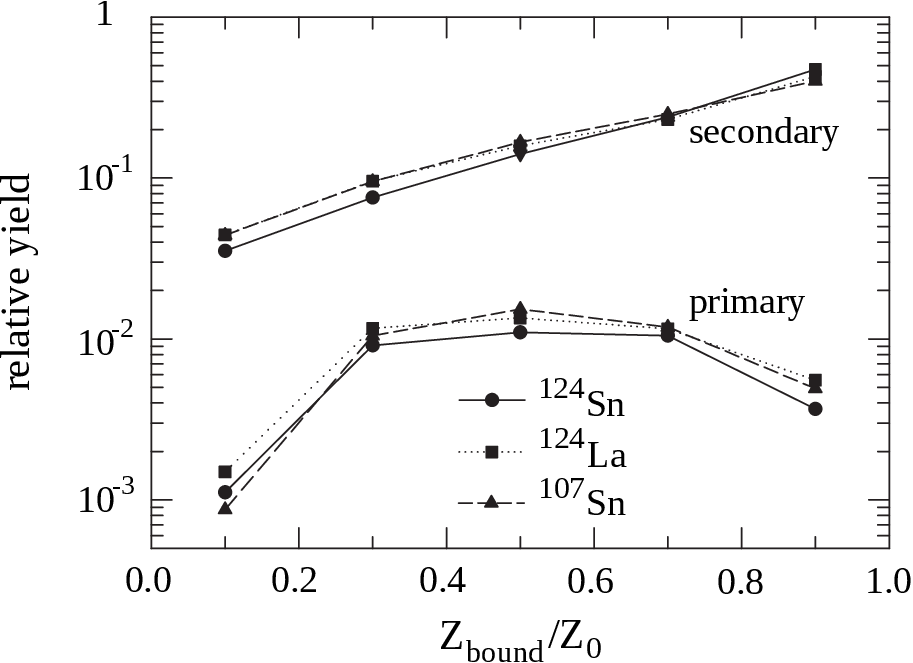}}
%\centerline{\includegraphics[width=8.5cm]{figures_smm/Fig17_primary-secondary.eps}}
\caption{\small{The SMM result for the relative yield of produced neutrons separated into the groups of neutrons from the primary disintegration 
of the ensemble of excited sources (primary) and from the secondary deexcitation of fragmentation products (secondary) 
as a function of $Z_{\rm bound}/Z_0$ for the three reactions as indicated. 
The three cases are each individually normalized. The lines connect the data symbols.}}
\label{fig:smm_hotcold}
\end{figure} %fig. 19
 
According to the calculations, the produced neutrons are mainly originating from the secondary decay of excited fragments as
illustrated in Fig.~\ref{fig:smm_hotcold}. The relative intensities of neutrons created 
in the primary partitioning of the excited sources are very small with $Z_{\rm bound}$-integrated strengths 
between 3.5\% and 4.5\% of the total yields. Their rise-and-fall behavior with a broad maximum centered at 
$Z_{\rm bound}/Z_0 = 0.5$ is similar to that of intermediate-mass fragments.

The yield of neutrons from secondary emissions is clearly dominant but the predicted $\zbound$ dependence is only
partly in agreement with the experimental data. In particular, at large $Z_{\rm bound}$, the neutron yields are overestimated.
One of the possible explanations for this discrepancy is related to the fact that the ensembles of sources were
adjusted to reproduce the observed fragment yields and correlations. The fragment yields are low at both ends of the 
$Z_{\rm bound}$ spectrum with the effect that the ensemble parameters are only poorly constrained. A closer inspection of the
predicted fragment yields shows that, e.g., the production of heavy residues with $Z \ge 40$ in the bin of largest $Z_{\rm bound}$ is
overestimated by a factor of perhaps 2 (Fig.~\ref{fig:z124} and Figs.~4 and 10-12 in Ref.~\cite{ogul11}).
It may contribute significantly to the neutron yields.

A second argument, addressing the underprediction of neutrons at smaller $Z_{\rm bound}$, may be related to the crescent-shaped
geometry expected for the spectator systems emerging from collisions with small impact parameter. They may be only weakly connected and
disintegrate into several smaller entities before they equilibrate~\cite{nepeivoda22}. Neutrons loosely attached to these structures 
may not participate in the statistical processes forming the sources accounting for the small number of fragments that are observed. 
Also in that case, fragment yields alone are not sufficient for a realistic description of the more complicated structure of the spectator system.
One may, therefore, conclude that not only the neutron multiplicity at small impact parameter is underpredicted for that reason 
but also the relative strength of the primary group of neutrons (Fig.~\ref{fig:smm_hotcold}). There is no doubt, however, that secondary
processes dominate at intermediate to large impact parameters. 

For the reasons just discussed, the bins of large and small $\zbound$ are not included in the evaluation of the temperatures characterizing the 
source of neutrons modeled with the SMM. The distributions of all neutrons integrated over the interval $10 \le \zbound \le 40$ exhibit 
temperature values $T = 4.5, 4.4,$ and 4.7~MeV for the fragmentations of $\sn$, $\snrad$, and $\la$, respectively. 
They are obtained by fitting the calculated momentum distributions in one dimension as, e.g., $p_x$ over the
interval $-100 \le p_x \le$ +100~MeV/$c$ with Gaussian functions. Choosing larger intervals leads to slightly higher temperatures, 
indicating that the distributions are not purely thermal and that several mechanisms may contribute. For $-150 \le p_x \le$ +150~MeV/$c$,
the temperatures are higher by $\approx 0.3$~MeV and for $\pm 200$~MeV/$c$ higher by $\approx 0.5$~MeV than those obtained with the $\pm 100$~MeV/$c$
interval.

The weak dependence on the projectile is similar as in the experiment. The average value $T = 4.5$~MeV is between 0.5 and 0.7 MeV higher 
than the experimental temperatures displayed in Fig.~\ref{fig:smmtemp} and smaller than the microcanonical breakup temperature by about 1.5~MeV. 
It represents an average temperature as expected for an evaporation cascade produced by excited projectile fragments with initial 
temperatures close to the microcanonical temperatures. Extending the Gaussian fits over larger momentum intervals seems to increase the weight of
sources with higher temperatures, including the small group of neutrons produced in the primary breakup (Fig.~\ref{fig:smm_hotcold}).

\section{Discussion}
\label{sec:dis}

\subsection{Bevalac data}
\label{sec:bevalac}

Among the Bevalac data reported by the group of Madey {\it et al.}, results of measurements performed at small angles 
for the reactions Ne on Pb at 390 and 790 MeV/nucleon~\cite{madey85} and for Nb on Nb and Au on Au at 800 MeV/nucleon~\cite{madey88} 
are well suited for a comparison with the present data.
%At the Bevalac, Madey {\it et al.} have systematically studied neutron emission in projectile fragmentation at relativistic
%energies and the dynamical properties of the identified sources~\cite{madey81,madey85,baldwin92,madey88,htun99}. 
%Measurements performed at small angles for the reactions Ne on Pb at 390 and 790 MeV/nucleon~\cite{madey85} 
%and for Nb on Nb and Au on Au at 800 MeV/nucleon~\cite{madey88} are well suited for a comparison with the present results. 
Plastic detectors of 10 cm thickness with a width of 25~cm and a length of about 1~m were used, 
so that the $0^{\circ}$ detectors positioned at 14.3~m from the target in the experiments with Nb and Au beams covered 
angles up to $2.0^{\circ}$ with respect to the beam direction. 
In the experiments with Ne beams, the flight path was 8.0~m and the 1-m-long detector at $0^{\circ}$ thus extended up to $3.6^{\circ}$.
In the neutron spectra measured at these forward angles, 
a ``striking peak" was discovered and found to be essentially independent of the projectile energy. 
The measured widths in the projectile frame were translated into source temperatures between 3.2 and 3.6~MeV and found to increase 
with decreasing impact parameter in the reactions with heavy projectiles~\cite{madey88}.

\begin{figure} [tbh]
\centerline{\includegraphics[width=8.0cm]{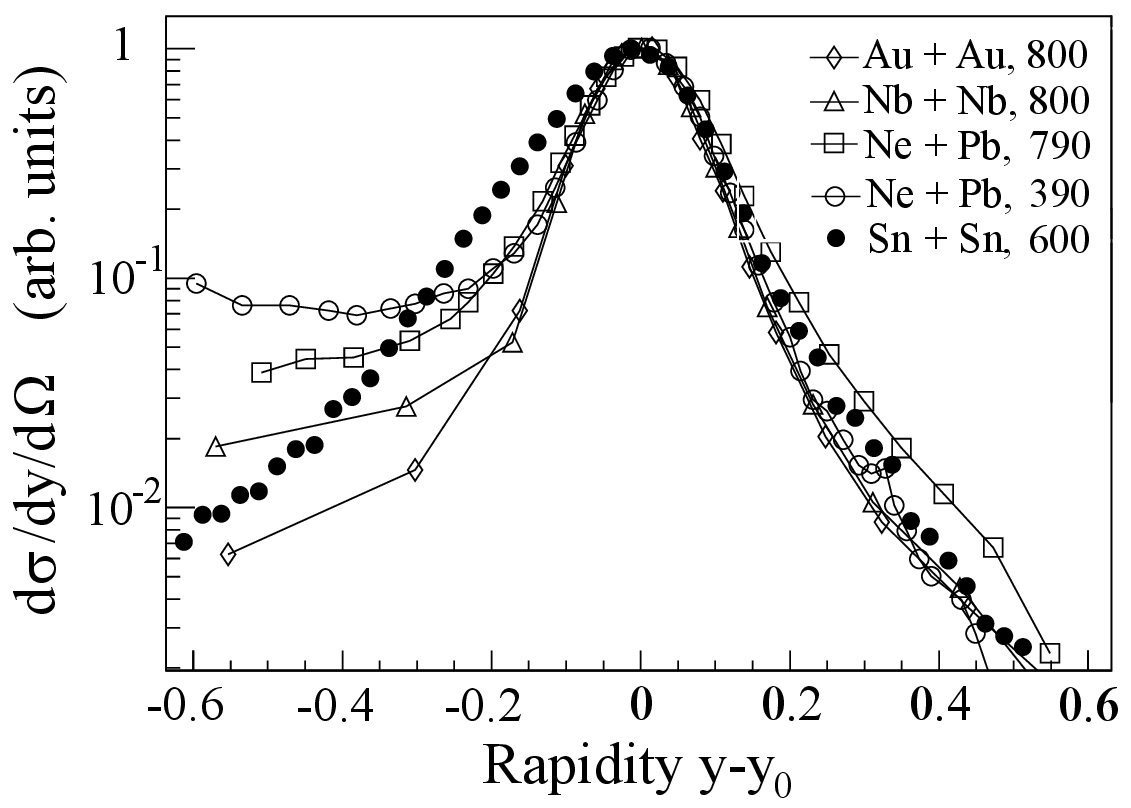}}
%\centerline{\includegraphics[width=8.5cm]{figures_wieloch/madey_new_v5.eps}}
\caption{\small{Rapidity distributions in the projectile frame ($y_0$ is the projectile rapidity) of neutrons at 
$\theta_{\rm lab} \le 2^{\circ}$ from the present $^{124}$Sn + Sn fragmentation reaction (filled circles) in comparison 
with several reactions studied at the Bevalac~\protect\cite{madey85,madey88}. The distributions are normalized with respect to their
maxima. The numbers in the legend indicate the projectile energies in MeV/nucleon.
}}
\label{fig:source_beva}
\end{figure} %fig. 20  

These findings are in very good agreement with the results obtained in the present study (Figs.~\ref{fig:2deg}--\ref{fig:temp_mult}), 
indicating a universal property of the spectator decay at relativistic energies. 
Rapidity spectra in the moving frame for the four reactions studied at the Bevalac are compared with the $^{124}$Sn + Sn case 
in Fig.~\ref{fig:source_beva}. With a normalization at the peak heights, the agreement is indeed impressive. 
The fall-off of the observed intensities toward larger rapidities is essentially identical for all five cases. 
On the low-energy side, at rapidities
$y - y_0 \le -0.2$, the different intensities of the Bevalac data presumably reflect different experimental 
conditions of the four experiments. In the present case of $^{124}$Sn, the peak is less symmetric and the slope toward smaller 
rapidity less steep. 

Slopes extending toward midrapidity are expected for light particles at small angles in projectile fragmentations. This was, for example, 
shown for $Z=2$ particles from $^{197}$Au reactions at 1000 MeV/nucleon~\cite{schuett96}. 
The slopes that are observed should depend strongly on the angular coverage~\cite{yordanov05}.
The ALADIN spectrometer used in these experiments has an acceptance of $\pm 10.2^{\circ}$ in the horizontal and $\pm 4.5^{\circ}$ 
in the vertical direction (Sec.~\ref{sec:s254}). In the Berkeley experiments, the acceptance was much more centered at $0^{\circ}$, 
thus apparently enhancing the detection of beam velocity neutrons with symmetric rapidity distributions. 
The slightly higher tail towards mid-rapidty observed with the coverage up to $3.6^{\circ}$ in the experiments with Ne beams 
would support this interpretation (Fig.~\ref{fig:source_beva}). 

\begin{figure}
\centerline{\includegraphics[width=7.7cm]{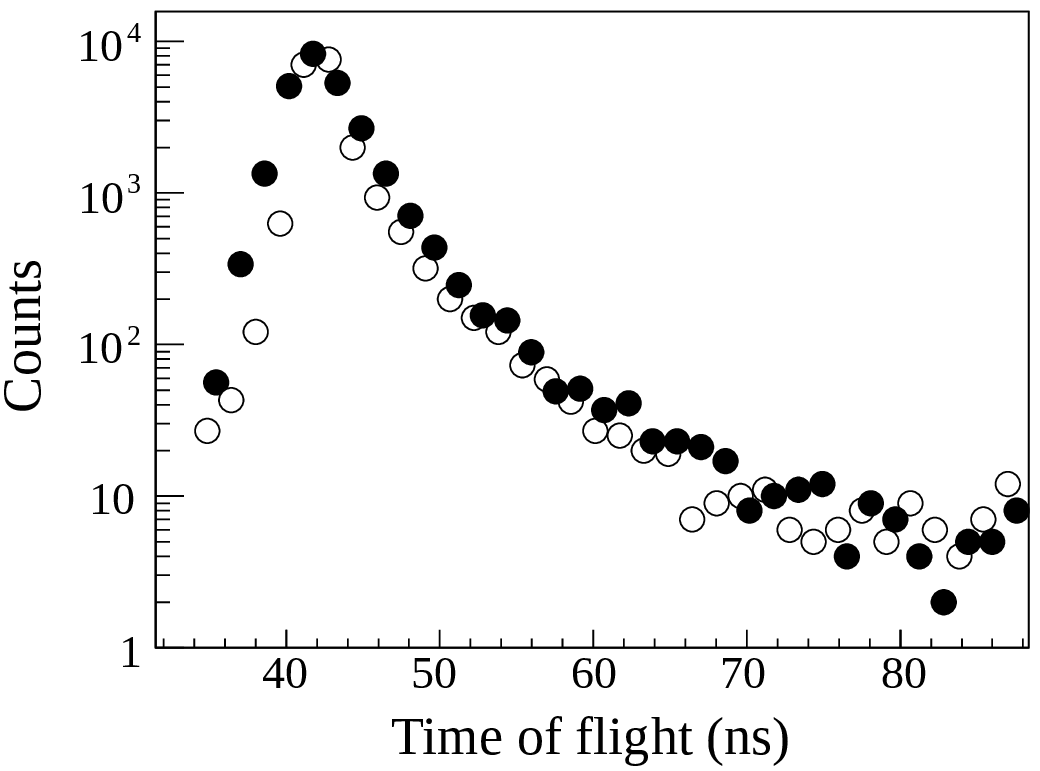}}
%\centerline{\includegraphics[width=8cm]{figures_wieloch/s254vss107_v4_new.eps}}
\caption{\small{Neutron time-of-flight spectra obtained in the present experiment (filled circles) 
and in experiment S107 with deuteron beams of 600 MeV/nucleon on a Pb target (open circles). 
The S107 distribution has been aligned in time and normalized at the peak. 
}}
\label{fig:piotr_tof}
\end{figure}

An alternative explanation for the modified slope appeared when the present data were compared with neutron time-of-flight spectra 
from the S107 test experiment which exhibit similar tails (Fig.~\ref{fig:piotr_tof}). In this experiment, 
performed for investigating properties of LAND~\cite{piotr_nima,boretzky03,gsi1992}, beam-velocity neutrons 
were generated by Coulomb and diffractive dissociation of deuterons in Pb targets. The distance to LAND
amounted to about 9.1~m which leads to slightly shorter flight times than in the present experiment but still permits 
a comparison of the peak structures. 
The tail to longer flight times recorded in S107 is not compatible with approximate symmetry in rapidity and thus not 
expected to originate from the breakup process.
%nor from sources of background in these calibration experiments.

     %fig. 21

The adopted explanation assumes that neutrons whose primary hits remain undetected produce delayed secondary hits after traveling through 
the detector for a certain time and possibly with reduced velocity. These hits if interpreted as the first hit of a neutron event
in the detector will cause a shift to lower velocity for this event. Because half of the detector volume consists of iron the probability 
for neutron interactions not producing sufficient light for being recorded is not negligible. 
It is thus not excluded that the present spectra at rapidities below the peak rapidity are affected by the described experimental effect and 
thus not necessarily in contradiction to the universality visible also here in the Berkeley data (Fig.~\ref{fig:source_beva}).

\subsection{Yields and temperatures}
\label{sec:yieldtemp}

The measured rapidity distributions have wider-than-Gaussian tails (Fig.~\ref{fig:source_beva}). Interpreting these shapes, 
Madey {\it et al.} have included additional Gaussians of larger widths to improve their description~\cite{madey85,madey88}. 
Several choices were found equally successful, leaving this procedure somewhat arbitrary. One of the choices, however, 
included a Gaussian of width $\sigma \approx 110~\MeV/c$ with an integrated strength comparable to that of the narrower peak. 
From the larger widths a momentum close to the Fermi momentum was obtained from the formula $P = \sqrt{5} \sigma$ 
of the Goldhaber model~\cite{gold74}. 

It is assumed in this model that the product momenta in fast fragmentation processes may reflect the nucleonic Fermi motion 
within the colliding nuclei. 
The resulting momentum distributions are indistinguishable from that of a thermalized system with high temperatures of the 
order of 9 MeV. Indications for such processes were found experimentally, although with momentum distributions corresponding
to significantly higher temperatures. Proton spectra measured for 
$^{197}$Au + $^{197}$Au collisions at 1000 MeV/nucleon exhibit two components with an evaporation bump superimposed on flatter 
tails with extracted 
temperatures $T \approx 6$ and $\approx 25$~MeV for the two components, respectively~\cite{odeh00}. 
Also the transverse momenta of light fragments in spectator decays measured in ALADIN 
experiments were found compatible with the Goldhaber model, the extracted temperatures $T \approx 15$~MeV being somewhat
closer to the expected value~\cite{schuett96}. 

Comparable transverse-momentum widths for the fragmentation of $^{197}$Au and
$^{208}$Pb projectiles at similar as well as much higher incident energies were reported by Chance {\it et al.}~\cite{chance01} 
and H\"{u}ntrup {\it et al.}~\cite{huentrup01}. The deduced momentum widths exceed the expectation according to the Goldhaber 
model by factors up to about 2. Possible explanations for these excess widths were presented by Bauer~\cite{bauer95}. 
Neutron components with comparable widths are difficult to identify in the 
present experiment because of the limited acceptance in transverse momentum. However, the measured rapidity distributions do not
exclude their existence (Fig.~\ref{fig:2deg}).

The SMM does not contain equivalent mechanisms as part of the fragmentation process. In the model, a common temperature 
for all degrees of freedom of a given configuration is assumed. 
The present results for neutron emissions at forward angles  
thus indicates that, at medium and large $Z_{\rm bound}$, the partitioning of the spectator sources into particles and 
fragments and their subsequent decay is 
%well described on the basis of the assumed equipartition and adequate deexcitation models for the final reaction stages. 
%Besides the excellent results for the production and correlations of fragments, also the yields of neutrons %and the temperatures 
sufficient to satisfactorily describe the multiplicities and temperatures observed for the low-temperature neutron component. 

The observed difference of measured and calculated neutron multiplicities at smaller $Z_{\rm bound}$
was discussed in the previous section. It seems to indicate that %for these more violent collisions, 
%the mechanisms of neutron production are not exhaustively described with the statistical breakup of the selected ensemble of sources. 
the decay of the ensemble of excited sources with properties adjusted to reproduce the fragmentation data
does not fully represent the production of
spectator neutrons in these more violent collisions.

In a dynamical study based on the IQMD transport 
model, neutron multiplicities in qualitative agreement with the data measured for $^{124}$Sn and $^{124}$La fragmentations 
were obtained~\cite{kumar12}. However, the location of the calculated neutron distributions at rapidities around 90\% of the
projectile rapidity, corresponding to $p_z = 1057$~MeV/$c$ or $E_{\rm lab}=478$~MeV, is not consistent with the experimental observations
(Figs~\ref{fig:source} and \ref{fig:2deg}).
Presumably, secondary emissions from excited projectile fragments were not explicitly considered.

Combining the measured and calculated information and ignoring for this purpose the experimental uncertainty, 
an approximate neutron balance may be attempted. For mid-peripheral collisions 
with $Z_{\rm bound}$ in the interval 20 to 30, the neutron multplicity measured for the spectator source is $M_n = 11$ in the case 
of $^{124}$Sn fragmentations (Fig.~\ref{fig:temp_mult}). Following Pietrzak {\it et al.}~\cite{pietrzak20}, the size of the spectator system 
corresponding 
to $Z_{\rm bound} = 25$ is $Z_{\rm spect} \approx 30$. In the participant-spectator picture, this leaves 20 charges from 
each of the two nuclei 
for the participant source. Applying again the relative abundance of nucleons and light charged particles measured 
by FOPI (cf. Sec.~\ref{sec:react}), 
one obtains a composition with 14 free protons, 26 protons and 29 neutrons bound in light charged particles, and 30 free neutrons in this source. 
Adding the 11 neutrons emitted by the projectile spectator (Fig.~\ref{fig:temp_mult}) and assuming $\approx 10$ neutrons bound in multifragmentation 
products not accounted for in the model of Bertini~\cite{bertini} yields 51 neutrons, a value close to the prediction shown in Fig.~\ref{fig:bertini}.
The neutrons emitted by the target spectator are assumed to be below the threshold chosen in the calculations. 
It allows the conclusion that the flat neutron yields underneath the observed peaks, 
possibly with momenta reflecting the internal Fermi motion according to Goldhaber~\cite{gold74}, are part of the participant source 
formed in the early stage of the reaction. The fact that this stage is not included in the SMM scenario, may also partly explain 
the underprediction of the neutron multiplicities at small $Z_{\rm bound}$ (Fig.~\ref{fig:smmnmult}).

\section{Summary}
\label{sec:con}

The production of neutrons at forward angles from the fragmentation and subsequent decay of the projectile spectators in reactions 
of $^{124}$Sn, $^{107}$Sn, and $^{124}$La projectiles with Sn targets at 600 MeV/nucleon was investigated with the LAND detector positioned 
approximately 10 m downstream from the target of the ALADIN spectrometer. The identification and localization of neutrons with this detector 
followed the method presented in Ref.~\cite{piotr_nima}. As reported there and discussed in the experimental sections, the multiplicity 
of neutrons in the present case of multi-neutron events cannot be determined better than with uncertainties of up to 20\%. 
Because of the loss of neutrons between the target and the LAND detector of the order of 10\%, the reported absolute
neutron cross sections and multiplicities are possibly too low by an amount of that order.

A thermal source of neutrons traveling with nearly exactly beam velocity was identified. The temperatures between 2 and 5 MeV extracted 
from transverse momentum distributions measured with LAND do not indicate any visible dependence on the isotopic composition of the projectile. 
Their average value is in good agreement with the results obtained at the Bevalac and supports the invariance with projectile mass 
and energy observed there for incident energies between 390 and 800 MeV/nucleon~\cite{madey85,madey88}. The observed invariance is expected
within the more general concept of limiting fragmentation~\cite{benecke69}, presently being discussed up to the highest energies available 
for heavy-ion reactions (for recent references see, e.g., Ref.~\cite{wolschin21}). Regarding the yields of projectile fragments, it 
manifests itself in the universality observed for $^{197}$Au and $^{208}$Pb fragmentation at energies from several hundred MeV/nucleon
up to 158 GeV/nucleon~\cite{schuett96,cherry95,huentrup00}. 

The neutron multiplicities measured at small $Z_{\rm bound}$, i.e., small impact parameters, reflect the neutron-to-proton ratio $N/Z$ of the projectile. 
They are, on average, 30\% higher for the case of $^{124}$Sn than for the two neutron-poor projectiles, corresponding to the differences
of their $N/Z$ ratios. At large $Z_{\rm bound}$, i.e., for reactions 
at large impact parameter leading to predominantly excited residues, the neutron multiplicities grow faster than in proportion to the absolute 
neutron numbers of the three projectiles. The different multiplicities represent the only isotopic effect of the neutron production 
observed in the present experiment.

The suitability of the ALADIN-LAND setup for the present purpose was studied with the Geant4 program package, using a simplified but in crucial 
dimensions accurate model geometry. Calculations were performed for neutron and fragment beams starting from the location of the target and with 
various properties, including moving thermal sources with temperatures up to 6 MeV. The calculations confirm that neutrons starting toward 
the high-efficiency volume defined within the detector volume of LAND arrive there with high probability without suffering interactions on 
their flight paths. 
The level of background from scattered neutrons with original directions other than toward the high-efficiency volume is small, of the order of 5\%
according to the estimates discussed in Sec.~\ref{sec:geant4}. 
With this limitation, the setup was found to be appropriate for the intended measurements, in retrospect justifying its use in 
earlier experiments~\cite{poch95,odeh00}. 

The statistical multifragmentation model (SMM) was used to calculate the neutron emission expected from ensemble calculations with parameters that 
permitted the successful interpretation of the fragment yields and correlations reported in Refs.~\cite{ogul11,pietrzak20}. 
The predicted multiplicities for larger impact parameters are close to the observed values and correctly reflect the measured 
dependence on the isotopic 
composition of the projectile. The dependence on the choice made for the coefficient $\gamma$ of the symmetry term of the liquid-drop 
description of the produced fragments was found to be small and significant only for the fragmentation of $^{124}$Sn projectiles. In this case, 
the multiplicity of evaporated neutrons is slightly decreasing for smaller values of $\gamma$, consistent with the simultaneous increase of 
the production of neutron-rich fragments~\cite{ogul11}. The calculated temperatures with an average value $T = 4.5$~MeV are between 0.5 and 0.7 MeV higher 
than the experimental values. The nearly negligible dependence on the neutron content of the projectile is similar as observed in the experiment. 

The SMM description established in the fragmentation study performed for this experiment thus permits also very satisfactory predictions
for neutrons, their multiplicities and source temperatures, at least for medium and large impact parameters. 
The underprediction of the observed neutron multiplicities at smaller impact parameters seems to indicate
that the decay of the ensemble of excited sources 
can only partly account for the production of spectator neutrons at forward angles in these more violent collisions.

\begin{acknowledgments}

A.S.B. is grateful to FIAS Frankfurt, the GSI Helmholtzzentrum Darmstadt, and the IKP of Mainz
University for support and hospitality.
R.O. thanks TUBITAK-DFG cooperation for financial support.
C.Sf. acknowledges support by an Alexander-von-Humboldt fellowship.
This work was supported by the European Community under Contract No. HPRI-CT-1999-00001, by the 
French-German Collaboration Agreement between IN2P3 - DSM/CEA and GSI, 
by the Polish Ministry of Science and Higher Education under Contracts No. N202 160 32/4308 
(2007-2009) and No. 1 P03B 020 30 (2006-2009), and by the Swedish Research Council under Contract No. 2022-04248.

\end{acknowledgments}

\section{Appendix: Study of the detection efficiency with Geant4}
\label{sec:app}

An extended study of the experimental ALADIN-LAND setup with calculations within the Geant4 framework~\cite{geant4} was performed to document 
its suitability for measuring neutron emissions in spectator decay. The elements of the spectrometer contain significant amounts of iron and 
other materials that are not directly shadowing LAND from the target position but are potential sources of background of neutrons and other 
particles produced by scattering and inelastic processes of neutrons emitted from the target. The ALADIN magnet, at the same time, 
serves as an effective shield by absorbing neutrons from the mid-rapidity source, emitted at larger polar angles in more central collisions.

\begin{figure}
%\centerline{\includegraphics[width=6.5cm]{figures_hakan/scattergun_2.eps}}
\centerline{\includegraphics[width=6.5cm]{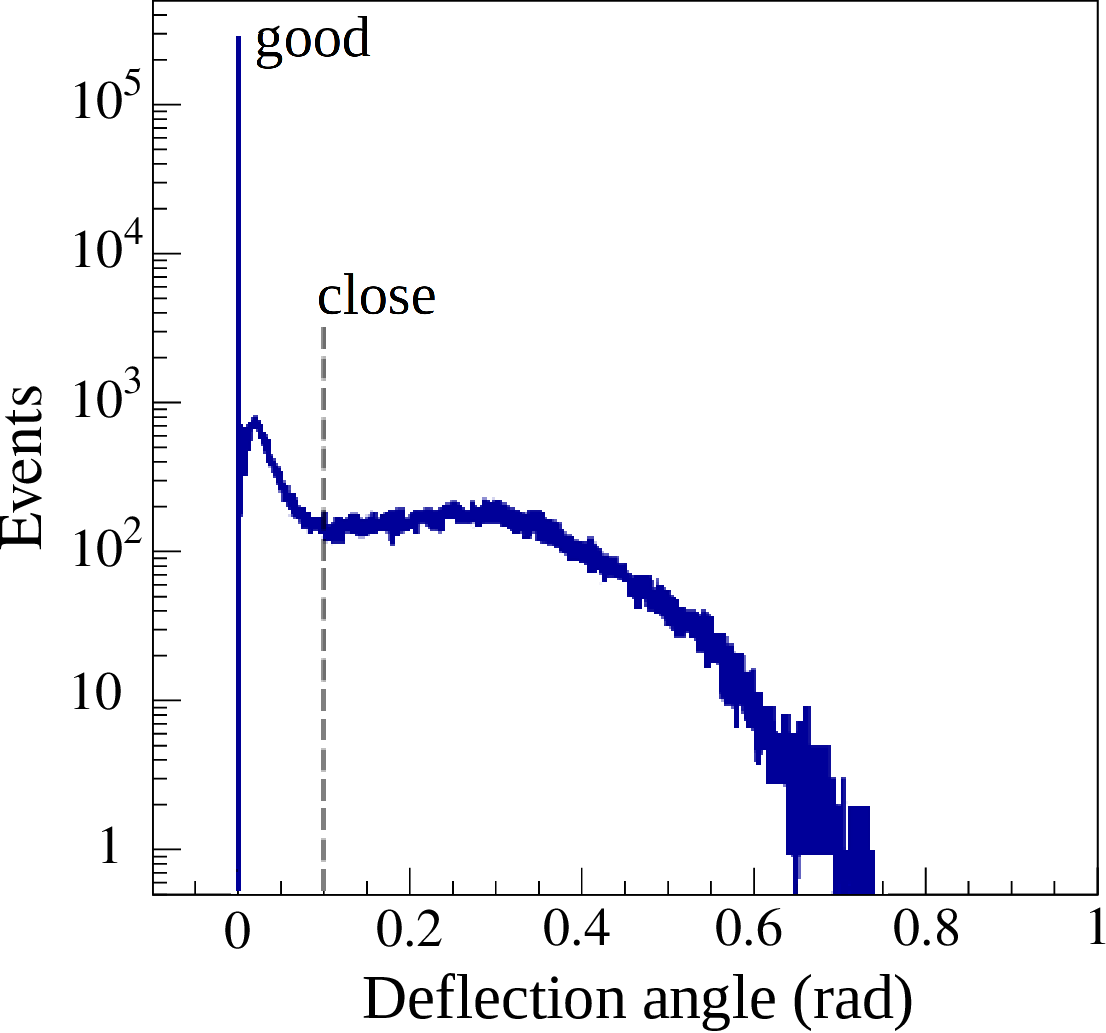}}
%\centerline{\includegraphics[width=6.5cm]{figures_wieloch/fig22.eps}}
\caption{\small{Distribution of displacements of 600-MeV neutrons from a source located at the 
target position and emitting isotropically within the solid angle given by $\theta_{\rm lab} \le 15^{\circ}$, 
expressed as apparent changes of the angle of emission as viewed from the target.
Deflection angles different from zero are caused by scattering processes experienced by the neutrons 
along their trajectories to the test plane. 
The labels ``good" and ``close", respectively, indicate that no scattering occurred and that, in the case of a scattering process, 
the final apparent direction is within a cone of $\le 0.1$~rad (5.7$^{\circ}$) with respect to the initial direction (dashed line).}}
\label{fig:scattergun}
\end{figure}     %fig. 22/A1

\begin{figure*}
%\centerline{\includegraphics[width=14cm]{figures_hakan/picturegun.eps}} 
\centerline{\includegraphics[width=16.0cm]{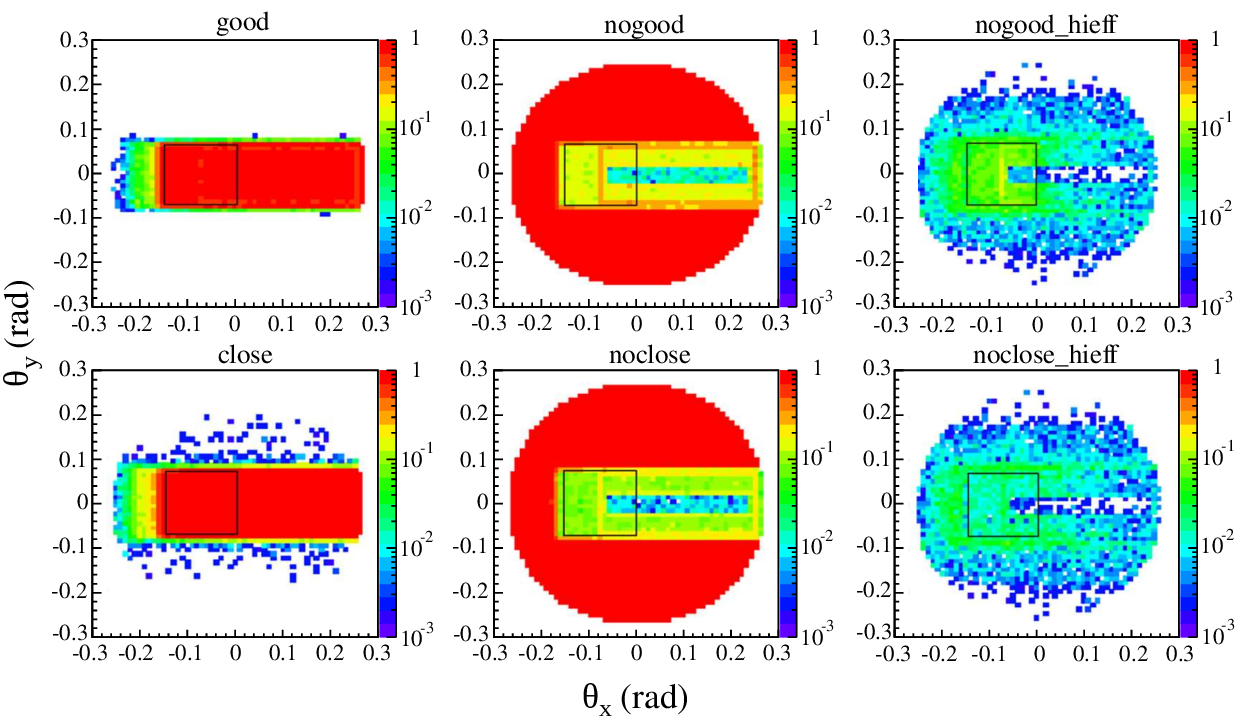}} 
%\centerline{\includegraphics[width=14.5cm]{figures_wieloch/fig23.eps}} 
\caption{\small{Original directions with respect to the beam direction of 600-MeV neutrons from a source located at the target position and emitting 
isotropically within the solid angle given by $\theta_{\rm lab} \le 15^{\circ}$, sorted according to the scatterings experienced along their 
trajectories to the test plane (see text for explanations). The black square represents the high-efficiency area plotted in the chosen coordinate 
system in which the negative $x$ direction viewed from the target position points to the right.
%The axis descriptions are the same for all six panels and identical to those of Fig.~\protect\ref{fig:alanogood}.
}}
\label{fig:picturegun}
\end{figure*}     %fig. 23/A2

The model of the experimental setup generated for this purpose was based on the measured dimensions and positions of the major 
components available with precisions on the level of $\pm 1$~mm. Minor components were represented with their geometries simplified, 
inhomogeneous surfaces smoothed, and technical details omitted, all under the condition that the total amount of material is preserved. 
The model was divided into the substructures ALADIN magnet, the vacuum and detector chambers, and the time-of-flight wall. 
The floor of and the air in the experimental vault was included as well. For evaluating the resulting effects, a test plane perpendicular to 
the primary beam direction at the location of the front plane of LAND was added to the model. Within this plane, a high-efficiency area was 
defined with the lateral dimensions of the high-efficiency volume (Sec.~\ref{sec:higheff}).

The first set of calculations was aimed at determining the scattering probabilities as a function of the emission angle from the target position. 
Pencil beams of neutrons with energies between 400 and 800 MeV were used for homogeneously scanning a 
cone of polar angles $\theta_{\rm lab} \le 15^{\circ}$. 
Their trajectories and those of produced secondary particles were followed and, if they were not absorbed earlier, 
registered with their positions on the test plane. Examples of the results obtained for 600-MeV neutrons are given in the following. 
They are different from those given in Sec.~\ref{sec:geant4} because the homogeneous solid-angle coverage chosen here is different from that
produced by moving thermal sources (see below).

More than half of the neutrons emitted into the $15^{\circ}$ cone are absorbed by the magnet and other parts of the setup and only about
37\% arrive at the test plane according to the calculations.
The displacements at the test plane caused by scattering processes are presented as angles of deflection with respect to the original direction
of emission as viewed from the target position. The resulting distribution for $10^6$ neutrons emitted within the defined cone 
is shown in Fig.~\ref{fig:scattergun}. The label ``good" in the figure indicates that no scattering has occurred. It is the case for 
approximately 70\% of the number of neutrons reaching the test plane. Small deflections form a narrow bump contained within 0.05 rad, followed
by a wider distribution with a broad maximum close to 0.3 rad. The label ``close" includes processes causing deflections up to 0.1 rad ($5.7^{\circ}$),
primarily produced by interactions within the time-of-flight wall and the exit flange of the detector chamber (cf. Fig.~\ref{fig:scatter4MeV}).

The observed effects are presented in Fig.~\ref{fig:picturegun} in the form of probabilities for the absence of any scattering (``good"), 
the occurrence of small-angle scatterings (``close") and for their complements (``nogood" and ``noclose"). In addition, the latter two groups are also 
tested for whether the neutron finally passes through the high-efficiency area and, with this condition met, plotted in the right column. 
The color code representing the probabilities within the two-dimensional plane of original emission angles is logarithmic and the same for all six cases. 

\begin{figure}
%\centerline{\includegraphics[width=6.5cm]{figures_hakan/hitsalanogood.eps}}
\centerline{\includegraphics[width=7.0cm]{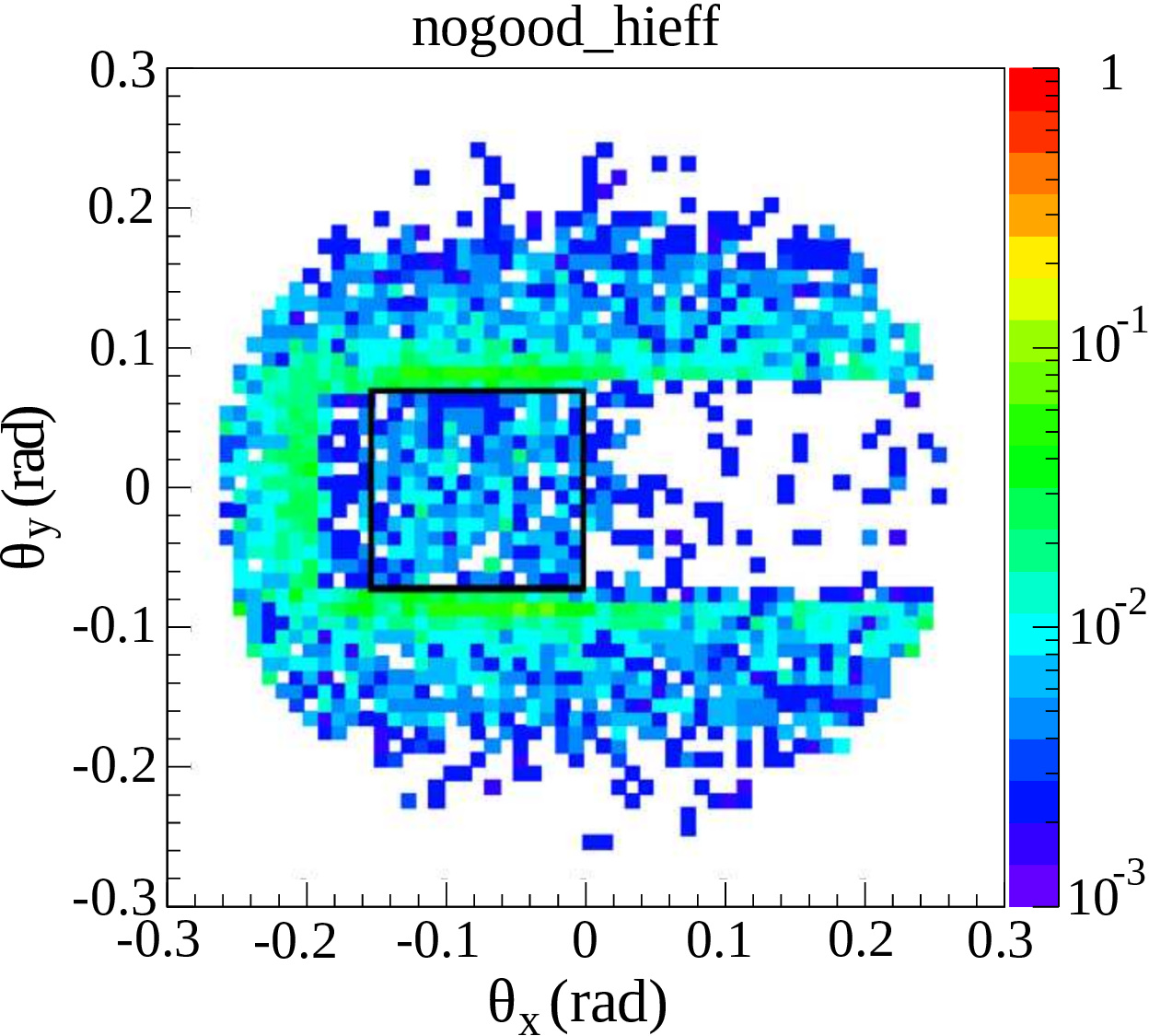}}
%\centerline{\includegraphics[width=6.5cm]{figures_wieloch/FIG_24.eps}}
\caption{\small{Probability distribution of the original directions with respect to the beam direction of 600-MeV neutrons from a source, 
located at the target position and emitting isotropically within the solid angle given by $\theta_{\rm lab} \le 15^{\circ}$, in a 
virtual setup including only the ALADIN magnet and the air within the vault. 
Events are recorded under the condition that a scattering occurs and the scattered neutron 
enters the high-efficiency area of the test plane represented by the black square. }}
\label{fig:alanogood} 
\end{figure}     %fig. 24/A3

The main observation is apparent in the left column of Fig.~\ref{fig:picturegun}. Neutrons can reach the test plane without scattering or 
with small deflections essentially only if they start toward the window defined by the poles, yoke and vacuum chamber of the ALADIN magnet. 
The high-efficiency area is contained within this window. However about 20\% of the neutrons emitted toward the high-efficiency window will 
suffer a scattering process (top-middle panel ``nogood"). About one half of them (probability 10\%) will still hit the high-efficiency 
area (top-right panel), most likely with only a minor deflection because the probability for being emitted toward the high-efficiency 
area and still hitting it after a deflection by more than 0.1 rad (``noclose") is on the level of 2\% or less (bottom-right panel 
of Fig.~\ref{fig:picturegun}). 
The narrow horizontal band with reduced scattering probability visible in the middle and right columns shows 
the effect of the thin window of 1-mm steel 
welded into the rear wall of the detector chamber (Fig.~\ref{fig:setup}). Viewed from the target position, its vertical width of 28 cm 
corresponds to angles of $\pm 22$~mrad.   

The modular structure of the setup as reproduced for the calculations made it possible to study the effects of individual components. 
As an example, results obtained by activating only the ALADIN magnet together with the floor of and the air within the vault are shown 
in Fig.~\ref{fig:alanogood}. 
It gives again the probabilities for scattered neutrons reaching the high-efficiency area as a function of their original directions 
but under the condition that only the magnet is present. The magnet cannot affect neutrons emitted towards the high-efficiency area. 
Scattering processes in the air of the vault are responsible for the intensity of 1\% or less visible within the black square representing 
the high-efficiency area. Neutrons emitted into other directions are scattered into the high-efficiency area with very similar probability. 
Bands of higher intensity ($\approx 5\%$) are
produced by neutrons emitted towards the upper and lower edges of the magnet poles and magnet chamber and scattered into the high-efficiency area.
The figure may be compared with the corresponding figure with the full setup shown in the top-right panel of Fig.~\ref{fig:picturegun}. 

More realistic emission profiles were modeled with thermal sources of temperatures up to 6 MeV moving with the projectile velocity. 
Examples of results for the 4-MeV source were already included in the main text (Sec.~\ref{sec:geant4}). 
The initial distribution of emissions (Fig.~\ref{fig:source4MeV}), 
the spectrum of deflections (Fig.~\ref{fig:scatter4MeV}), and the distribution of scattered neutrons entering the high-efficiency 
area (Fig.~\ref{fig:nogood4MeV}) are shown there. 
The essential results are very similar to those shown in Fig.~\ref{fig:picturegun}.
The scattering probabilities as a function of the initial directions are necessarily the same but the fewer emissions 
at larger polar angles reduce the overall background intensity, thus leading to the smaller values given in Sec.~\ref{sec:geant4} and
the narrower distribution of deflections shown in Fig.~\ref{fig:scatter4MeV}. 

A final set of calculations was performed with fragment beams emitted from the target and bent by the magnet into directions close to 
that of the deflected primary projectiles. The aim was to follow the history of fragments produced in the target and traveling toward 
the time-of-flight wall and exit flange of the detector chamber. Interactions with the light nuclei of the plastic time-of-flight detectors can be expected 
to produce mainly excited fragment residues, evaporating neutrons into narrow cones in their flight directions. 
Fragments with transverse momenta in vertical directions large enough for them to miss the thin exit window installed for 
the projectile beam may undergo more violent reactions with the thicker steel parts of the chamber.

For investigating the level of background to be expected from these processes, the Bertini model~\cite{bertini} was selected 
from the physics list of Geant4 (cf. Fig.~\ref{fig:bertini}). The majority of secondary neutrons originates from evaporation into 
a narrow cone of directions with respect to the direction of the emitting fragment and most likely from a position somewhere in the 
time-of-flight wall or exit flange of the detector chamber. It was found that the probability for reaching the high-efficiency volume of 
LAND is at most a few percent per fragment or $\alpha$ particle, and concentrated on the side of the deflected beam. With the measured $\alpha$ and 
IMF multiplicities, this will add up to not more than typically 10 \%, i.e. $\approx 0.1$ neutron per event. Extrapolation from 
the high-efficiency volume to the full phase space of the projectile source increases this value by a factor of roughly 3. It still 
permits the conclusion that secondary reactions of projectile fragments in the time-of-flight wall or the exit flange are not significantly enlarging 
the background. The latter consists mainly of neutrons produced in the target and undergoing 
scattering processes somewhere along their trajectories.

\end{document}